\documentclass[]{bytedance_seed}

\usepackage[toc,page,header]{appendix}

\usepackage{minitoc}
\usepackage{cleveref} 
\usepackage{subcaption}
\usepackage{booktabs}

\ifdefined\final
  \usepackage[disable]{todonotes}
\else
  \usepackage[textsize=tiny]{todonotes}
\fi
\usepackage{natbib}
\usepackage{latexsym}

\usepackage{url}
\usepackage{amssymb}
\usepackage[utf8]{inputenc}
\usepackage{microtype}
\usepackage{booktabs}
\usepackage{pifont} 
\usepackage{multirow}
\usepackage{makecell}
\usepackage{paralist}
\usepackage{xspace}
\usepackage{color}
\usepackage{xcolor}
\usepackage{colortbl}
\usepackage{adjustbox}
\usepackage{hyperref} 
\usepackage[edges]{forest}
\usepackage{tikz} 
\usepackage{caption}
\usepackage{amsfonts}

\hypersetup{
    colorlinks,
    linkcolor={blue!80!black},
    citecolor={blue!80!black},
}
\tikzset{
    root/.style =             {align=center, text width=1cm, rounded corners=3pt, line width=0.3mm, fill=gray!10, draw=gray!80, font=\small},
    demographic/.style =         {align=center, text width=1.8cm, rounded corners=3pt, line width=0.3mm, fill=blue!10, draw=blue!80, font=\footnotesize},
    demographic_work/.style =    {align=center, text width=10cm, rounded corners=3pt, line width=0.3mm, fill=blue!10, draw=blue!0, font=\footnotesize},
    character/.style =         {align=center, text width=1.8cm, rounded corners=3pt, line width=0.3mm, fill=red!10, draw=red!80, font=\footnotesize},
    character_work/.style =    {align=center, text width=10cm, rounded corners=3pt, line width=0.3mm, fill=red!10, draw=red!0, font=\footnotesize},
    personalization/.style =           {align=center, text width=1.8cm, rounded corners=3pt, line width=0.3mm, fill=cyan!10, draw=cyan!80, font=\footnotesize},
    personalization_work/.style =      {align=center, text width=10cm, rounded corners=3pt, line width=0.3mm, fill=cyan!10, draw=cyan!0, font=\footnotesize},
    risk/.style =         {align=center, text width=1.8cm, rounded corners=3pt, line width=0.3mm, fill=orange!10, draw=orange!80, font=\footnotesize},
    risk_work/.style =    {align=center, text width=10cm, rounded corners=3pt, line width=0.3mm, fill=orange!10, draw=orange!0, font=\footnotesize},
}

\usepackage{CJK}

\title{SeedFold: Scaling Biomolecular Structure Prediction}

\affiliation[1]{ByteDance Seed}

\contribution{Full author list in Contributions}

\abstract{
Highly accurate biomolecular structure prediction is a key component of developing biomolecular foundation models, and one of the most critical aspects of building foundation models is identifying the recipes for scaling the model. In this work, we present \texttt{SeedFold}, a folding model that successfully scales up the model capacity. Our contributions are threefold: first, we identify an effective width-scaling strategy for the Pairformer to increase representation capacity; second, we introduce a novel linear triangular attention that reduces computational complexity to enable efficient scaling; finally, we construct a large-scale distillation dataset to substantially enlarge the training set. Experiments on FoldBench show that \texttt{SeedFold} outperforms AlphaFold3 on most protein-related tasks.
}

\date{Dec 30, 2025}

\checkdata[Correspondence]{Quanquan Gu (\url{quanquan.gu@bytedance.com})}
\checkdata[Project Page]{\url{https://seedfold.github.io/}}

\begin{document}

\maketitle


\vspace{-30pt}
\begin{figure}[h]
    \centering
    \includegraphics[width=0.95\linewidth]{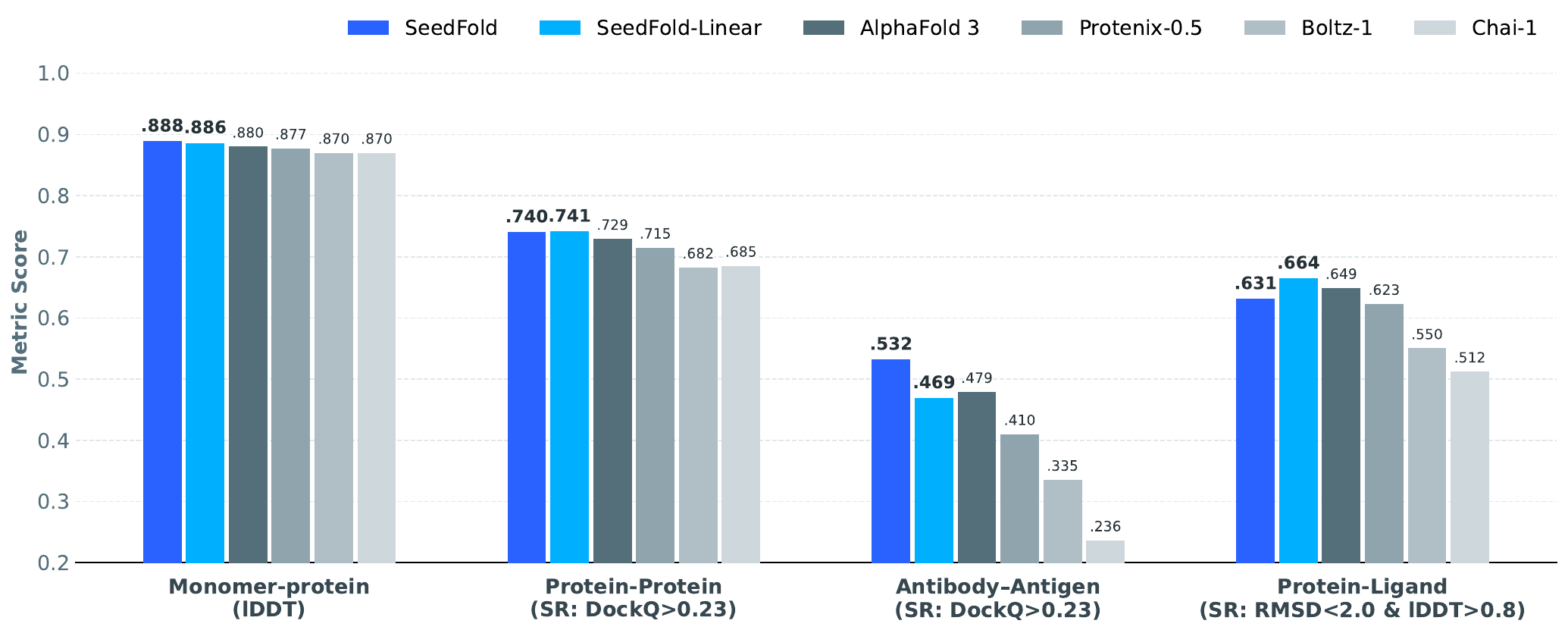}
    \caption{Overview of \texttt{SeedFold}. (i) We scale folding models from three perspectives. \textbf{Model}: scaling the Pairformer width to increase model capacity; \textbf{Architecture}: linear triangular attention reduces computational complexity while maintaining prediction quality; \textbf{Data}: large-scale distillation expands training data to $26.5$M samples. (ii) \texttt{SeedFold} denotes a 512-width model equipped with vanilla triangular attention, while \texttt{SeedFold-Linear} refers to a 384-width model with linear attention. (iii) \texttt{SeedFold} achieves state-of-the-art performance on FoldBench, surpassing AlphaFold3 and other open-source models on multiple tasks. }
    \label{fig:front}
\end{figure}

\newpage
\section{Introduction}

Determining protein structures is crucial for structural biology and drug discovery. Experimental techniques, including X-ray crystallography and cryo-electron microscopy, facilitate the construction of high-resolution protein models; however, these approaches are considerably time-consuming. AlphaFold2~\cite{jumper2021highly} revolutionized protein structure prediction, achieving experimental-level accuracy for protein monomers in 2021. AlphaFold3~\cite{abramson2024accurate} further unifies the modeling of proteins, DNA, RNA, and ligands at an atomic level. There has been a line of research works~\cite{ahdritz2024openfold,wohlwend2025boltz,bytedance2025protenix,qiao2024neuralplexer3,chai2024chai,team2025intfold} focusing on high accuracy structure prediciton at AlphaFold-level, which has made a great contribution to the community. 

Folding models harness prior knowledge and exhibit versatility across an extensive spectrum of applications, including structure generation~\cite{watson2023novo,krishna2024generalized}, binder design~\cite{pacesa2024bindcraft}, conformation sampling~\cite{monteiro2024high}, or even in turn benefits experimental methods~\cite{raghu2025multiscaleguidanceproteinstructure,jamali2024automated}. Therefore, building effective folding models is a critical step towards creating protein foundation models. Modern foundation models such as GPT-4~\cite{abramson2024accurate} exhibit superior artificial general intelligence, yet their architectural design does not differ substantially from that of their predecessor, GPT-2, except for scaling the model size to an exponentially larger magnitude~\cite{radford2019language}. This raises two research questions: \textit{is current model capacity sufficient for protein structure modeling}, and \textit{what is the best way to scale these models}?

Although a few research works have attempted to examine the scaling behavior of such folding models~\cite{qiao2024neuralplexer3,passaro2025boltz}, most recent folding models adhere to the basic configurations of AlphaFold. They have mainly focused on scaling the number of Pairformer layers, while we find the performance of folding models may be bounded by the hidden dimension, which bottlenecks their capacity. The recycling strategy of AlphaFold (up to $3$ at the training stage and $9$ at the inference stage) has already approximated the increase in the model depth, which has proven better at representation learning than its non-recursive counterpart~\cite{lan2019albert,yang2023looped}. In this study, we revisit the architecture of current folding models and investigate approaches to scaling the folding models with a more scalable architecture. 

\paragraph{Model Scaling} We identify the key factors governing the scaling of model size among three options: deepening the Pairformer module ($48 \to 96$), deepening the Structure module ($24 \to 48$), and widening the Pairformer module ($128 \to 256 \to 384 \to 512$). Experiments demonstrate that the module in folding models is sufficiently deep to support reasoning in the latent space, whereas the model capacity is primarily bottlenecked by the hidden dimension of the pairwise representation ($128$). Our observation aligns with DeepSeek-V3~\cite{liu2024deepseek} which contains $671$B parameters: while the number of layers is only $61$, the hidden size is increased to $7168$.
\paragraph{Linear Triangular Attention} Through inspecting each component of AlphaFold3, we can easily identify the computational bottleneck - triangular operations in the Pairformer. The complexity of triangular operations scales cubically with the length of proteins, which consumes substantial time and memory. We propose to draw upon modern techniques from large language models, where \texttt{softmax}-based triangular attention can be replaced by linear attention~\cite{sun2024liteformer,wu2025flashbias}, thereby reducing the complexity from cubic to quadratic. 

\paragraph{Large Scale Data Distillation} A large-scale dataset characterized by high quality and diversity serves as a key ingredient for modern deep learning models. However, the number of experimentally determined structures remains limited. The structure module of AlphaFold 3 (i.e., the Transformer) lacks the inductive biases inherent to AlphaFold 2, such as rotational and translational invariance. It may not generalize well when trained on insufficient amounts of data~\cite{dosovitskiy2020image}. To address this challenge, our solution is to construct a large-scale dataset derived from AlphaFold 2. It is well established that knowledge distillation from models with stronger regularization confers benefits to the learning process of models with weak regularization~\cite{gu2017non}.

We present \texttt{SeedFold}, a folding model that scales the Pairformer module and introduces a novel linear triangular attention mechanism to replace the computationally expensive vanilla triangular attention. We benchmark our model on FoldBench~\cite{xu2025benchmarking}. The results demonstrate that \texttt{SeedFold} surpasses existing open-source models, including Protenix-0.5~\cite{bytedance2025protenix}, Boltz-1~\cite{passaro2025boltz} and Chai-1~\cite{chai2024chai}, across all evaluation metrics. 
Notably, our vanilla and linear attention models display divergent learning behaviors: although both surpass AlphaFold3 on protein monomers and protein-protein complexes, their superior performance is task-specific. The vanilla model outperforms AlphaFold3 in antibody-antigen interactions, whereas the linear attention model performs well in protein-ligand interactions. This finding underscores the value of combining heterogeneous attention mechanisms to optimize model performance for targeted applications.

\section{Method}
In this section, we outline the scaling strategies from three perspectives. In Section \ref{sec:model_scaling}, we detail the scaling strategies of the folding models. In Section \ref{sec:linear_att}, we propose a novel triangular attention module as a drop-in replacement for the vanilla triangular attention module. In Section \ref{sec:data_scaling}, we present the dataset crawled for training \texttt{SeedFold}.

\begin{figure}[t]
    \centering
    \begin{subfigure}{\linewidth}
        \centering
        \includegraphics[width=0.85\linewidth]{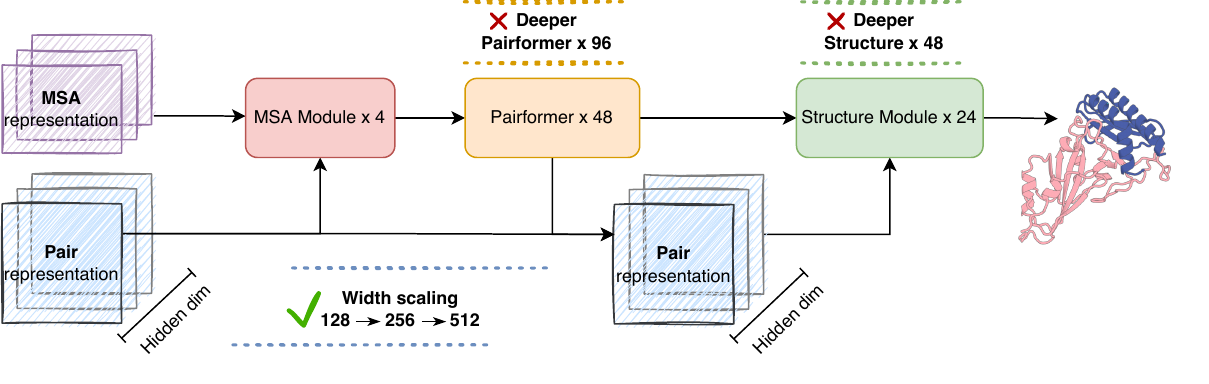}
        \caption{Model architecture and scaling strategies.}
        \label{fig:arch}
    \end{subfigure}
    \vspace{1em}
    \begin{subfigure}{\linewidth}
    \centering
        \includegraphics[width=0.8\linewidth]{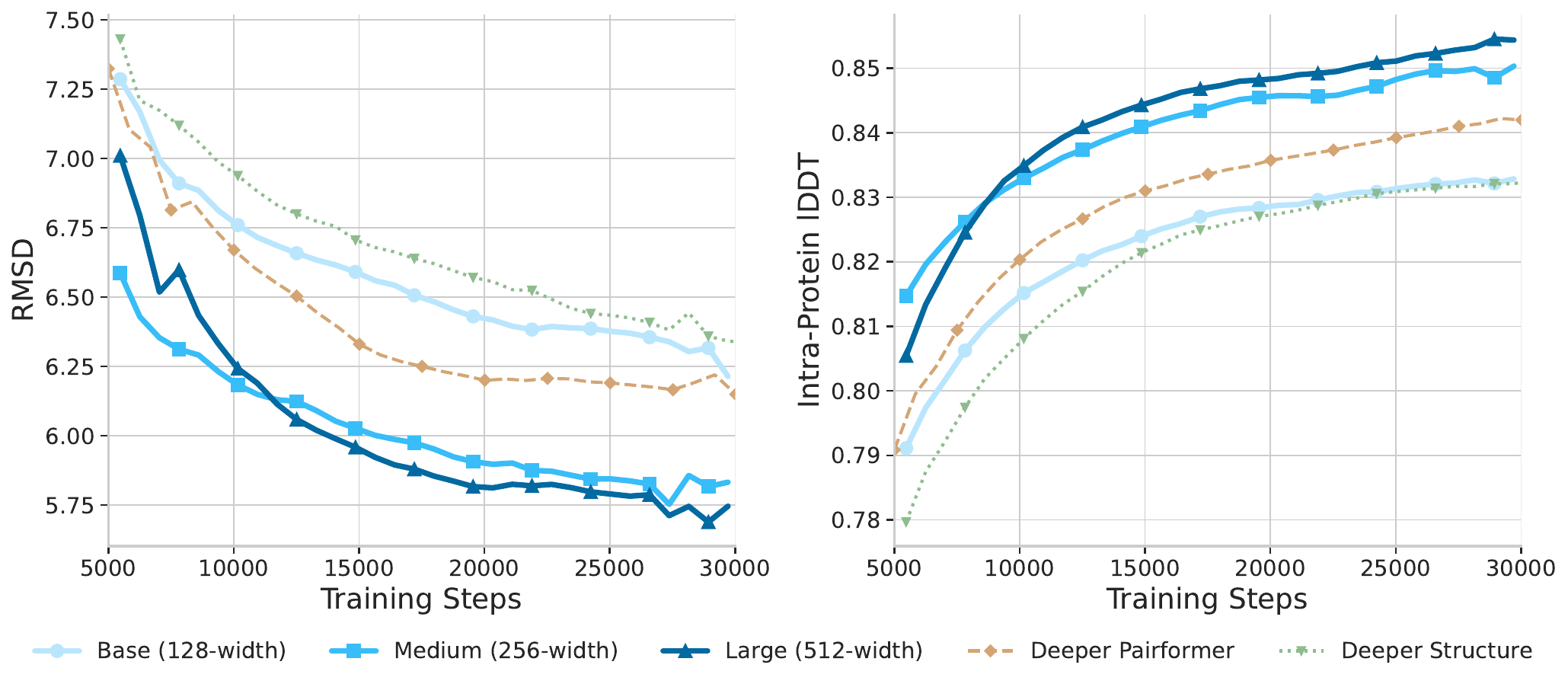}
        \caption{Comparsion of different scaling strategies.}
        \label{fig:scaling_data}
    \end{subfigure}
    \caption{Comparison of different scaling strategies. (a) Conceptual illustration of the scaling strategy.(b) Performance comparison where the left plot shows complex RMSD (lower is better) for global structural accuracy, and the right plot shows intra-protein lDDT (higher is better) for local structural quality. Width scaling consistently outperforms depth scaling. The transition from Small ($128$-width) to Medium ($256$-width) yields the largest gains in both global RMSD and local lDDT, while further scaling to Large ($512$-width) shows diminishing returns. Deeper trunk and deeper structure module provide marginal improvements compared to width scaling.}
    \label{fig:scaling_combined}
\end{figure}

\subsection{Model Scaling}
\label{sec:model_scaling}

The capacity of current folding models is primarily bottlenecked by the hidden dimension of the pair representation. While previous works have mainly focused on scaling the number of Pairformer layers~\cite{passaro2025boltz,bytedance2025protenix}, this work investigates scaling the hidden dimension, hypothesizing it to be more effective for representation learning. To explore the upper bound of model capacity, the model is systematically scaled up by increasing the pair representation dimension from $128$ to $512$, with a corresponding expansion in the MSA module to provide richer evolutionary features.

\paragraph{Model Architecture Overview}
Several strategies can be explored for model scaling. The folding model primarily consists of two modules: the trunk module and the structure module. The trunk module is mainly composed of the MSA module and the Pairformer module. The MSA module samples diverse MSAs through different recycles and extracts evolutionary features, which are then used to update the pair representation; conversely, the learned pair representation is also used to update the MSA hidden representation. The Pairformer module updates pair representations through pairwise triangular multiplication and triangular attention to capture inter-token interactions. The structure module takes the pair and single representations encoded by the trunk module as conditions to perform all-atom structure generation.

\begin{table}[t]
    \centering
    \caption{Model configurations at different scales. We explore scaling strategies for folding models by varying the trunk width (pair and MSA dimensions), trunk depth (Pairformer layers), and structure module depth. Training efficiency is measured on the same GPU type for fair comparison.}
    \label{tab:model_scaling}
    \begin{tabular}{lcccccc}
        \toprule
        \textbf{Configuration} & \textbf{\makecell{Pair Rep.\\Dim}} & \textbf{\makecell{MSA Rep.\\Dim}} & \textbf{\makecell{Pairformer\\Layers}} & \textbf{\makecell{Structure\\Layers}} & \textbf{\makecell{\#Params}} & \textbf{\makecell{Efficiency\\(iters/s)}} \\

        \midrule
        Base (128-width) & $128$ & $64$ & $48$ & $24$ & $432$M & $0.15$ \\
        Medium (256-width) & $256$ & $128$ & $48$ & $24$ & $533$M & $0.10$ \\
        Large (512-width) & $512$ & $256$ & $48$ & $24$ & $923$M & $0.06$ \\
        \midrule
        Deep Pairformer & $128$ & $64$ & $96$ & $24$ & $582$M & $0.10$ \\
        Deep Structure Module & $128$ & $64$ & $48$ & $48$ & $706$M & $0.10$ \\
        \bottomrule
    \end{tabular}
\end{table}


\paragraph{Scaling Strategy}
We construct three types of models to investigate different scaling strategies. Table~\ref{tab:model_scaling} presents the basic configurations of our scaling experiments. Figure~\ref{fig:scaling_combined} shows the results of different scaling strategies.

\begin{itemize}
    \item \textbf{Wider Trunk}: We scale the trunk width, i.e., the MSA module and the Pairformer width, as they work together to build rich pair representations. Specifically, we progressively increase the MSA hidden dimension from $64$ to $256$, while simultaneously increasing the dimension of pair representation from $128$ to $512$. The Pairformer parameter count scales from $147$M to $549$M, and the overall model size grows from $430$M to approximately $1$B. The corresponding training efficiency is shown in Table~\ref{tab:model_scaling}. 
    When scaling from Base ($128$-width) to Medium ($256$-width), we observe significant improvements across all metrics, including overall lDDT and RMSD that reflects global structural accuracy. When further scaling from Medium to Large ($512$-width), we adjust the optimization hyperparameters to stabilize training. The model still demonstrates consistent improvements on the test set, although the gains are less pronounced compared to the Base-to-Medium transition, suggesting diminishing returns at larger scales.
    
    \item \textbf{Deeper Trunk}: We increased the number of Pairformer layers from $48$ to $96$. This approach yielded a less significant performance gain compared to width scaling. This result is not surprising, as the use of $9$ recycling iterations with a $48$-layer network already creates a very deep effective architecture. Further increasing the physical depth is thus expected to offer diminishing returns. This observation is analogous to findings in large language models, where techniques like early exiting can skip later layers to accelerate inference with minimal impact on performance~\cite{elhoushi2024layerskip}.
    
    \item \textbf{Deeper Structure Module}: We increase the token transformer layers in the structure module from $24$ to $48$. This yields the lowest observed improvement among all scaling strategies. This is as expected, as the structure is largely determined by the pair representation, while the structure module functions more as a ``translator'' to convert a pair representation into coordinates~\cite{lin2023evolutionary}, which is a relatively easier task.
\end{itemize}

These results suggest that width scaling of the Pairformer is the most effective strategy for improving folding model performance. The pair representation dimension appears to be the critical bottleneck: increasing it directly enhances the model's capacity to encode complex pairwise interactions. In contrast, depth scaling of either the trunk or structure module provides diminishing returns, indicating that the representation capacity, rather than the number of processing steps, is the primary limiting factor.

\subsection{Linear Triangular Attention}
\label{sec:linear_att}

A scalable foundational component is essential for scaling folding models. At present, the main computational bottleneck of folding models occurs in the triangular operations, in particular within \texttt{TriangularAttention}. The module updates the pair representation $\mathbf{Z} \in \mathbb{R}^{n\times n \times d}$ in a manner similar to ``row-wise/column-wise attention with bias''. Denote the $i$-th row of $\mathbf{Z}$ as $\mathbf{Z}_i \in \mathbb{R}^{n\times d}$, the triangular attention on the $i$-th row can be computed by:
\begin{align}
\mathbf{Q}_i, \mathbf{K}_i, \mathbf{V}_i & = \mathtt{Linear}(\mathbf{Z}_i) \in \mathbb{R}^{n\times d} \\
\mathbf{B} & = \mathtt{Linear}(\mathbf{Z}) \in \mathbb{R}^{n\times n} \\
\mathtt{TriAtt}(\mathbf{Z}_i) & = \mathtt{softmax}\left(\mathbf{Q}_i\mathbf{K}_i^T+\mathbf{B}\right)\mathbf{V}_i,
\end{align}
where we omit the $\sqrt{d}$ in the attention for simplicity. Computing the attention scores for each row and column requires materializing a large matrix $(\mathbf{Q}_i\mathbf{K}_i^T+\mathbf{B}) \in \mathbf{R}^{n\times n}$, a cost that becomes particularly evident when the dimension of the row is taken into account and the complexity increases to  $\mathcal{O}(n^3d)$.

\begin{figure}[t]
    \centering
    \begin{subfigure}[t]{0.4\linewidth}
        \centering
        \includegraphics[width=\linewidth]{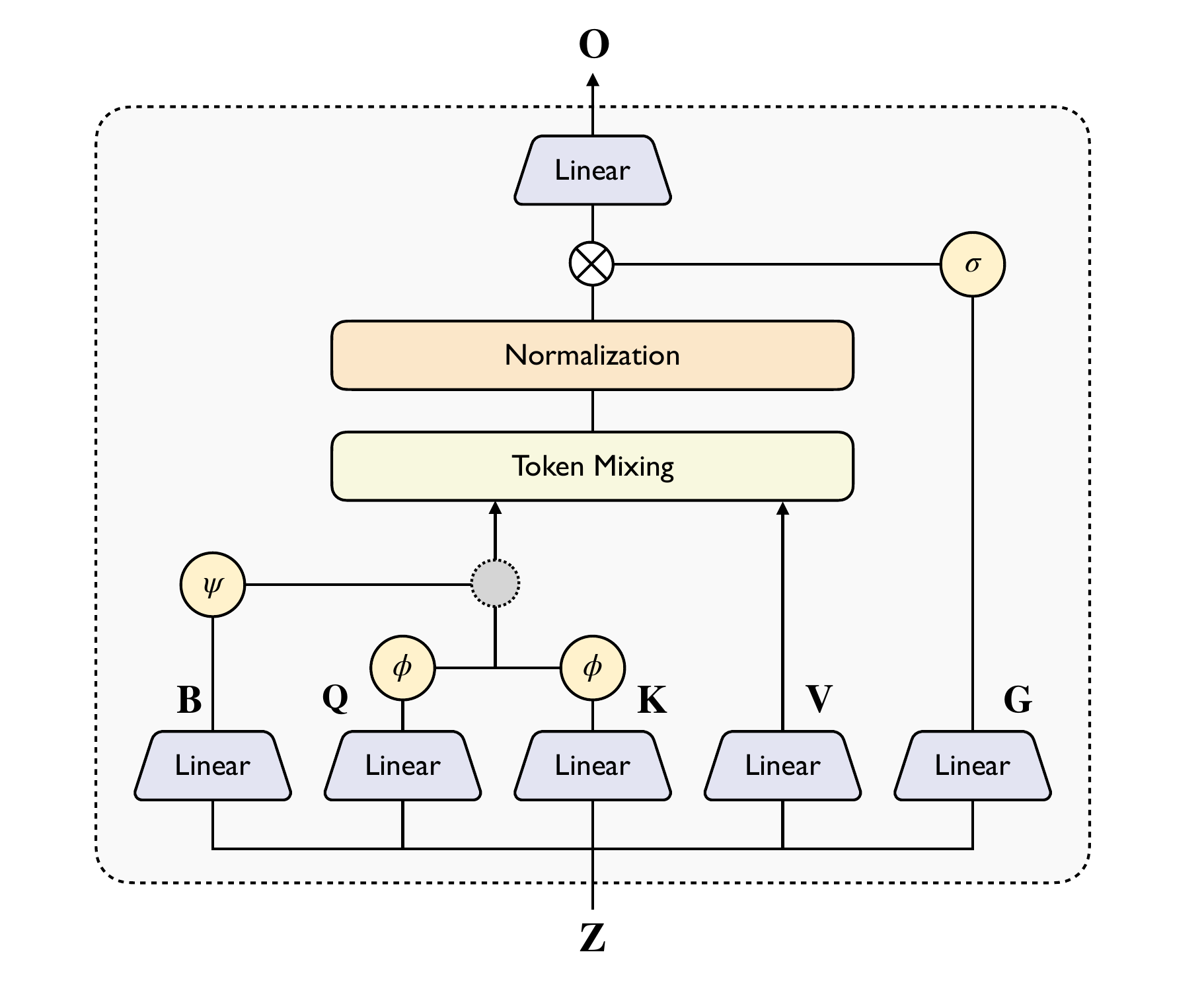}
        \caption{}
        \label{fig:linear_attention}
    \end{subfigure}
    \hfill
    \begin{subfigure}[t]{0.55\linewidth}
        \centering
        \includegraphics[width=\linewidth]{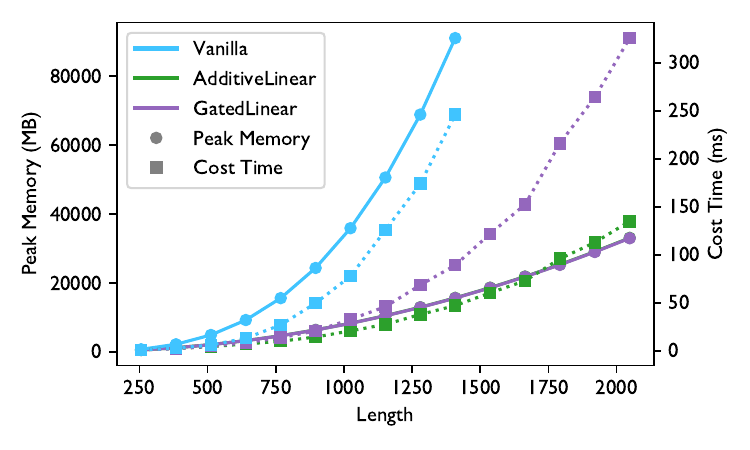}
        \caption{}
        \label{fig:bench_kernel}
    \end{subfigure}
    \caption{The \texttt{LinearTriangularAttention} module. (a) The architecture of the linear attention module. (b) The peak memory usage (MB) and time cost (ms) of different attention modules. Two linear attention mechanisms have similar peak memory usages, which overlap in the figure.}
\end{figure}

We aim to reduce complexity by introducing linear attention, which is a modern memory-saving technique widely adopted in large language models~\cite{minimax2025minimax01scalingfoundationmodels,kimiteam2025kimilinearexpressiveefficient}. The theoretical background of linear attention lies in the use of well-established random feature maps $\phi(\cdot)$ to approximate the ``dot-then-exponentiate'' function via the inner product $\mathtt{exp}(\mathbf{x}, \mathbf{y})\approx\langle\phi(\mathbf{x}), \phi(\mathbf{y})\rangle$ ~\cite{rahimi2007random,peng2021randomfeatureattention}. In practice, this is further simplified by directly substituting the \texttt{softmax} kernel with a simple non-linear feature map (e.g., $\mathtt{relu(\cdot)}$, $1+\mathtt{elu(\cdot)}$, $\mathtt{silu}(\cdot)$), where the attention operation $\mathtt{softmax}(\mathbf{Q}_i\mathbf{K}_i^T)\mathbf{V}_i$ is replaced by $\phi(\mathbf{Q}_i)\phi(\mathbf{K}_i)^T\mathbf{V}_i$. Such a reformulation offers a significant benefit, as it enables the application of the right product trick to reduce computational complexity. By first computing the right matrix multiplication, the computational cost can be reduced from $\mathcal{O}(n^2d)$ to  $\mathcal{O}(nd^2)$:
\begin{align}
\underbrace{\phi(\mathbf{Q}_i)\phi(\mathbf{K}_i)^T}_{\mathcal{O}(n^2d)}\mathbf{V}_i
\rightarrow \phi(\mathbf{Q}_i)\underbrace{\phi(\mathbf{K}_i)^T\mathbf{V}_i}_{\mathcal{O}(nd^2)}.
\end{align}
However, the original formulation of linear attention does not incorporate the bias term, which is essential for folding models to perform geometric reasoning in 3D space. Since $\mathbf{Z}_{ij}$ stores the coupling between the $i$-th and $j$-th tokens, $\mathbf{Q}_{ij}\mathbf{K}_{ik}$ quantifies the correlation between the $(i,j)$-th coupling and the $(i,k)$-th coupling. AlphaFold enhances the attention mechanism through the introduction of an additive bias, yielding a triangular form that also accounts for the $(j,k)$-th coupling: $\mathbf{Q}_{ij}\mathbf{K}_{ik}+\mathbf{B}_{jk}$.
We propose to convert the linear attention into a triangular form where the attention scores can be reweighted.
\begin{align}
    \mathtt{LinearTriAtt}(\mathbf{Z}_i) = \left(\phi(\mathbf{Q}_i)\phi(\mathbf{K}_i^T)\Box\psi(\mathbf{B})\right)\mathbf{V}_i,
\end{align}
where $\phi$ and $\psi$ are two feature maps and $\Box$ denotes an undefined operation. We present two choices for the boxed operation $\Box$: (i) an additive operation $+$ that up-weights or down-weights the attention score, similar to AlphaFold. (ii) a multiplication operation $\times$ that functions as a gating mechanism to control the information flow.

\paragraph{Additive Linear Triangular Attention} The additive operation inherits all the advantages of the vanilla version, which holds the potential to facilitate the adoption of well-established designs from modern linear attention. 
\begin{align}
    \mathtt{AdditiveLinearTriAtt}(\mathbf{Z}_i) & = \left(\phi(\mathbf{Q}_i)\phi(\mathbf{K}_i^T)+\psi(\mathbf{B})\right)\mathbf{V}_i \\
    & = \phi(\mathbf{Q}_i)\underbrace{\left(\phi(\mathbf{K}_i^T)\mathbf{V}_i\right)}_{\text{linearized}} +\underbrace{\psi(\mathbf{B})\mathbf{V}_i}_{\text{amortized}}.
\end{align}
It is worth noting that although we highlight the advantage of linear attention in avoiding the materialization of the $\mathbb{R}^{n\times n}$ attention matrix, the bias term in this variant is still structured as $\mathbb{R}^{n\times n}$. However, since triangular attention is performed for each row, its complexity is actually $\mathcal{O}(n\times n^2)$, where the first $n$ is the number of rows. With regard to the bias term, it is shared across all rows; that is, the memory usage can be amortized and does not grow with $n$. In our implementation, we choose $\phi=\mathtt{relu}$ and $\psi=\mathtt{relu}$.

\paragraph{Gated Linear Triangular Attention} The gated variant is more intuitive for controlling the information flow, i.e., we adopt a gating function $\psi:\mathbf{B}_{jk}\mapsto[0,1]$ to decide how much correlation exists between $\mathbf{Q}_{ij}$ and $\mathbf{K}_{ik}$.
\begin{align}
    \mathtt{GatedLinearTriAtt}(\mathbf{Z}_i) = \left(\phi(\mathbf{Q}_i)\phi(\mathbf{K}_i^T)\odot\psi(\mathbf{B})\right)\mathbf{V}_i, 
\end{align}
where $\odot$ denotes the pointwise matrix multiplication, which serves a role analogous to that of the (causal) attention mask in large language models. Unfortunately, the $\odot$ operation disrupts the matrix chain product, which means we cannot adopt the right product trick. To mitigate this issue, we develop a tiled version of linear attention which is optimized for reducing the memory footprint on CUDA devices~\cite{dao2022flashattentionfastmemoryefficientexact,Atkinson2025latkins,Geiger2025NVIDIA}. We refer the readers to Appendix~\ref{appendix:triton} for detailed information. In our implementation, we choose $\phi=\mathtt{relu}$ and $\psi=\mathtt{sigmoid}$.

Putting everything together, we developed two variants of linear triangular attention to mix information between tokens. The architecture of this module is illustrated in Figure~\ref{fig:linear_attention}. We further perform normalization and gating on the output following Lightning Attention, which has proven its effectiveness in industry-level large language models~\cite{qin2024lightning, qin2024transnormerllm, minimax2025minimax01scalingfoundationmodels}, i.e., $\mathtt{Linear}(\sigma(\mathtt{Linear}(\mathbf{Z}_i))\odot\mathtt{LayerNorm}(\mathtt{LinearTriAtt}(\mathbf{Z}_i)))$. Figure \ref{fig:bench_kernel} illustrates the peak memory usage and time cost of different attention modules, including the vanilla triangular attention and the two linear attention mechanisms proposed in this paper.



\subsection{Large Scale Data Distillation}
\label{sec:data_scaling}

In this section, we describe the data curation process for \texttt{SeedFold}. Detailed statistics are provided in Table~\ref{tab:data_stats}. We conducted large-scale data distillation to expand the training dataset to $26.5$M, which is $147$ times the size of the experimental dataset ($0.18$M). The motivation behind large-scale distillation stems from the paradigm shift from AlphaFold2 to AlphaFold3, where the inductive bias within the structure module is eliminated. AlphaFold2 employs the Invariant Permutation Attention (IPA) module to conduct reasoning in Euclidean space. In AlphaFold3, this module has been replaced with a general-purpose Transformer. Nevertheless, Transformers require large amounts of data and fail to generalize well when the dataset is small~\cite{dosovitskiy2020image}.

\begin{table}[t]
    \centering
    \caption{Dataset Statistics. The training set can be divided into two parts: the experimental dataset and the distillation dataset. We sample from these two types of datasets with equal probability.}
    \label{tab:data_stats}
    \begin{tabular}{cccc}
        \toprule
        \textbf{Dataset} & \textbf{Type} & \textbf{\#Samples} & \textbf{Weight} \\
        \midrule
        PDB & Experimental & $180,540$ & $0.50$ \\
        AFDB & Distillation & $3,326,991$ & $0.08$ \\
        Mgnify & Distillation & $23,075,211$ & $0.42$ \\
        \bottomrule
    \end{tabular}
\end{table}

The training set we used can be divided into three parts: 
\begin{itemize}
    \item \textbf{Protein Structures} We utilized Boltz's data processing pipelines~\cite{passaro2025boltz}, with PDB structures~\cite{berman2000protein} included up to the training date cutoff of 2021-09-30. Each chain and interface is assigned to a cluster. During training, samples are weighted by both molecular type and cluster size to ensure balanced representation.
    \item \textbf{AFDB Dataset} We obtained $3.3$M structures from the AFDB database, an opensource dataset which is generated by AlphaFold2~\cite{varadi2022alphafold}. We first performed chain sequence clustering with a minimum sequence identity of $0.5$, and then filtered out structures with a pLDDT score below $0.8$. We employed the AFDB dataset as the source for short monomers. Specifically, we first determine whether to sample monomers shorter than $200$ with a probability of $0.08$, and then perform uniform sampling of a chain within the specified length range.
    \item \textbf{Mgnify Dataset} We constructed a large-scale distillation dataset based on the Mgnify dataset~\cite{mitchell2020mgnify,evans2021protein}. We first filtered out sequences with fewer than $200$ residues. The sequences were clustered using MMSeqs~\cite{steinegger2017mmseqs2} with a minimum sequence identity of $0.3$. We employed \texttt{colabfold\_search}~\cite{mirdita2022colabfold} to generate multiple sequence alignments (MSAs) from the \texttt{uniref30\_db} and \texttt{colabfold\_envdb} sequence databases. We used OpenFold~\cite{ahdritz2024openfold} to infer high-quality protein structures using AlphaFold2's official weights. 
\end{itemize}

Here we briefly discuss the differences between the AFDB dataset and the MGnify dataset from three perspectives: the source, the diversity, and the length. (i) AFDB comprises AlphaFold-predicted 3D structures of proteins derived from UniProt sequences~\cite{uniprot2019uniprot}, whereas MGnify is a metagenomic dataset containing microbiome data from environmental microbial populations~\cite{mitchell2020mgnify}. (ii) From the perspective of model training, sample diversity is of greater concern. Our examination of the MGnify dataset revealed that among its $23$ million samples, only $2$ million could be assigned to clusters in the AFDB, thereby highlighting the sequence diversity. (iii) Figure~\ref{fig:data_length_dist} (in Appendix \ref{appendix:data}) presents the length distribution of the two distillation datasets, with the AFDB dataset having a median length of $95$ and the MGnify dataset a median length of $435$. This difference indicates that the Mgnify dataset may be beneficial for modeling long proteins.

\section{Training}
\label{sec:training}

\paragraph{Two-Stage Training Strategy}
We employ a two-stage training strategy to balance computational efficiency and model performance:

\begin{itemize}
    \item \textbf{Stage 1 (Small Crop Size)}: We train on crops of $384$ tokens with a diffusion batch size of $64$ for $60$k iterations. This stage allows for faster iteration and efficient exploration of the loss landscape.
    \item \textbf{Stage 2 (Large Crop Size)}: We increase the crop size to $640$ tokens and reduce the diffusion batch size to $32$, training for an additional $40$k iterations. This stage improves the model's ability to handle longer sequences and complex structures.
\end{itemize}

\paragraph{Training Configuration}
All models are trained with a batch size of $256$ using the AdamW optimizer~\cite{loshchilov2017decoupled}. We adopt a step decay learning rate schedule with a maximum learning rate of $0.0018$ for our base model. During training, we randomly drop MSA with a probability of $10\%$ and sample monomer distillation data with a ratio of $50\%$. 

\paragraph{Precision} Since the Structure Module operates in the coordinate space, while the Pairformer learns the latent space, their training and inference require different levels of precision. We use \texttt{bfloat16} for the MSA Module and the Pairformer, and consistently ensure that the Structure Module uses \texttt{float32}. Our experimental practice demonstrates that this approach achieves a balanced tradeoff between precision and computational speed. It is worth noting that if the Structure Module is run with \texttt{bfloat16}, local distance metrics (lDDT) will decrease significantly, whereas global metrics (DockQ) will remain relatively unchanged. This phenomenon is explicable because full precision exerts a considerably larger impact on local distances.

\paragraph{Training Stability}
Scaling the model width introduces training instabilities. When the Pairformer width exceeds $256$, we observe gradient norm explosions and loss collapse in early training. We adopt the following techniques to stabilize training:

\begin{itemize}
    \item \textbf{Extended Warmup}: We extend the warmup period from $1000$ to $3000$ to facilitate the gradual adaptation of large parameter matrices.
    \item \textbf{Reduced Learning Rate}: We use a smaller learning rate ($0.001$) for larger models ($512$-width). This aligns with common practice in large language models, where larger models require smaller learning rates to stabilize training~\cite{touvron2023llama}.
\end{itemize}

These techniques enabled the successful training of the large model to convergence. In this paper, we mainly present two models: 

\begin{itemize}
    \item \textbf{SeedFold}: a 512-width model equipped with vanilla triangular attention. Training folding models using a vanilla attention-based model yields greater stability, compared with its linear counterparts.
    \item \textbf{SeedFold-Linear}: a 384-width model integrated with \texttt{GatedLinearTriAtt}. We chose \texttt{GatedLinearTriAtt} due to its superior performance on DNA/RNA sequences, compared to \texttt{AdditiveLinearTriAtt}. We did not continue scaling the linear attention-based models due to resource limitations and potential convergence issues, and we will leave this for future work.
\end{itemize}
\section{Experimental Results}
\label{sec:main_results}

We conduct comprehensive evaluation on FoldBench~\cite{xu2025benchmarking}, a standardized benchmark covering diverse biomolecular structure prediction tasks. We compare \texttt{SeedFold} against state-of-the-art methods including AlphaFold 3~\cite{abramson2024accurate}, Boltz-1~\cite{wohlwend2025boltz}, Protenix~\cite{bytedance2025protenix}\footnote{\texttt{https://github.com/bytedance/Protenix}, version 0.6.1.}, and Chai-1~\cite{chai2024chai}. The performance of AlphaFold 3, Boltz-1, Chai-1 are cited directly from FoldBench. 

\begin{table}[t]
    \centering
    \caption{Main results on FoldBench. Success rates are reported for interface tasks (DockQ $\geq 0.23$ for protein interfaces,  lRMSD $< 2$\AA~and lDDT-PLI $> 0.8$ for ligands). For all metrics, higher is better. The results of AlphaFold3, Boltz-1 and Chai-1 are retrieved from the FoldBench paper \cite{xu2025benchmarking}. We re-evaluated Protenix-0.5, a newer version than the one employed in FoldBench. For monomers, the underline symbol $\_$ denotes that results with higher precision are not provided in FoldBench.}
    \label{tab:main_results}
    \begin{tabular}{lcccccccc}
        \toprule
        \textbf{Model} & \makecell{\textbf{Monomer}\\\textbf{(lDDT)}} & \makecell{\textbf{Prot-Prot}\\\textbf{(DockQ)}} & \makecell{\textbf{Ab-Ag}\\\textbf{(DockQ)}} & \makecell{\textbf{Prot-Lig}\\\textbf{(SR\%)}} &
        \makecell{\textbf{Prot-RNA}\\\textbf{(DockQ\%)}} &
        \makecell{\textbf{Prot-DNA}\\\textbf{(DockQ\%)}} &
        \makecell{\textbf{DNA}\\\textbf{(lDDT)}} &
        \makecell{\textbf{RNA}\\\textbf{(lDDT)}}
        \\
        \midrule
        AlphaFold 3 & $0.88\_$ & $72.93\%$ & $47.90\%$ & $64.90\%$ & $62.32\%$ & $\mathbf{79.18\%}$ & $0.61\_$ & $\mathbf{0.53\_}$ \\
        Boltz-1 & $0.87\_$ & $68.25\%$ & $33.54\%$ & $55.04\%$ & $56.90\%$ & $70.97\%$ & $0.44\_$ & $0.34\_$ \\
        Chai-1 & $0.87\_$ & $68.53\%$ & $23.64\%$ & $51.23\%$ & $50.91\%$ & $69.97\%$ & $0.49\_$ & $0.46\_$\\
        Protenix-0.5 & $0.8773$ & $71.50\%$ & $41.00\%$ & $62.30\%$ & $50.70\%$ & $71.38\%$ & $\mathbf{0.6249}$ & $0.4930$\\
        \midrule
        \textbf{SeedFold} & $\mathbf{0.8889}$ & $74.03\%$ & $\mathbf{53.21}\%$ & $63.12\%$ & $\mathbf{65.31}\%$ & $72.60\%$ & $0.5897$ & $0.4990$ \\
        \textbf{SeedFold-Linear} & $0.8861$ & $\mathbf{74.14\%}$ & $46.91\%$ & $\mathbf{66.48\%}$ & $61.80\%$ & $76.00$ & $0.5965$ & $0.4610$ \\
        \bottomrule
    \end{tabular}
\end{table}

\subsection{Benchmark}


\paragraph{Experimental Setup}
FoldBench consists of $1,522$ biological assemblies spanning nine prediction tasks, with structures deposited after 2023-01-13. The benchmark includes: (i) \textbf{Monomers}: $334$ protein, $14$ RNA, and $15$ DNA monomers; (ii) \textbf{Interfaces}: $279$ protein-protein, $172$ antibody-antigen, $558$ protein-ligand, $70$ protein-RNA, $330$ protein-DNA, and $51$ protein-peptide interfaces. All targets are filtered for low homology to training sets to ensure fair evaluation. Following FoldBench, we adopt the $5\times5$ sampling strategy ($5$ random seeds $\times$ $5$ diffusion samples) with $10$ recycling steps. The best prediction is selected using the model's ranking score.

\begin{figure}[t]
    \centering
    \begin{minipage}[t]{0.32\textwidth}
        \centering
        \includegraphics[width=\linewidth]{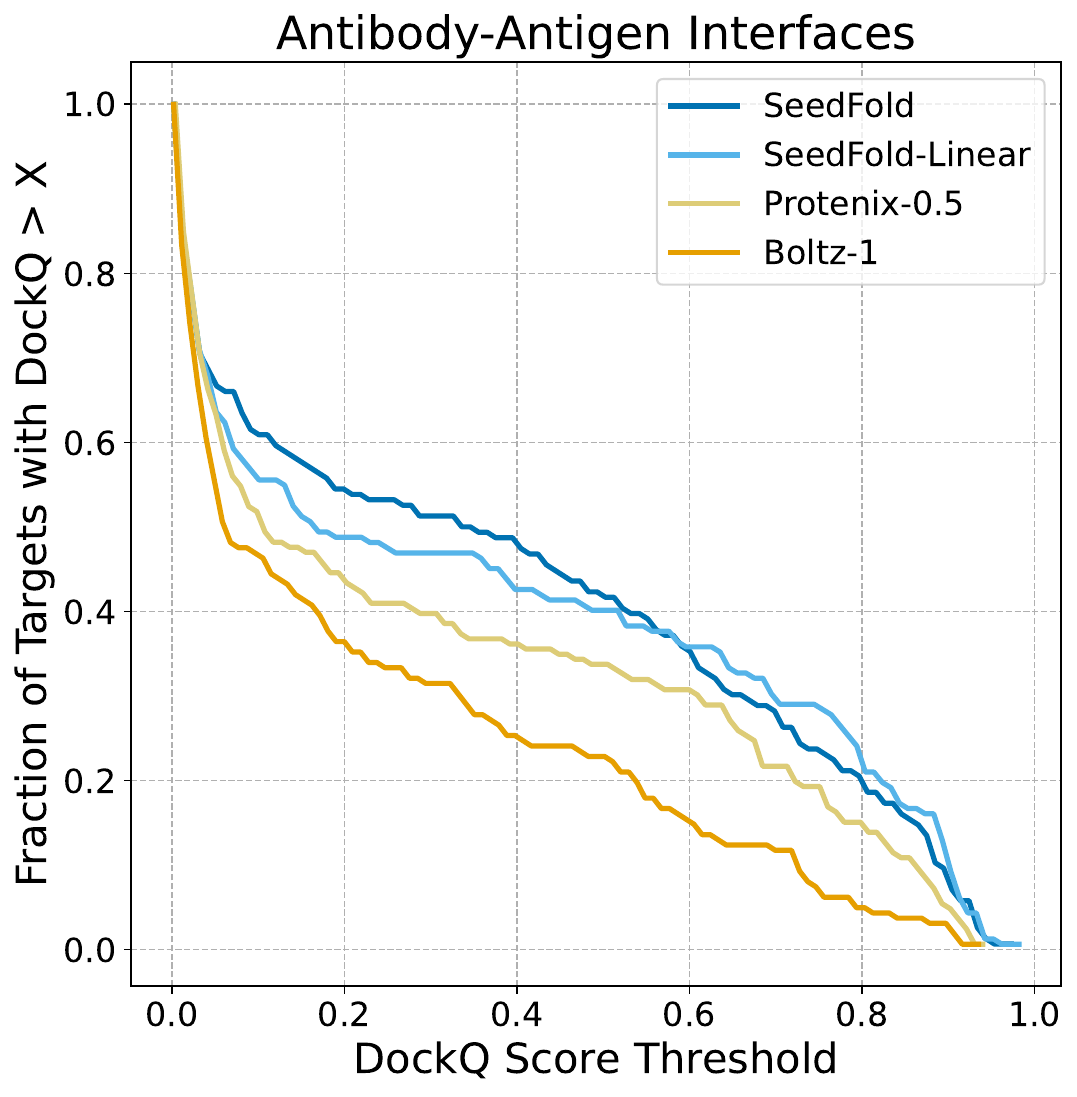}
    \end{minipage}
    \hfill
    \begin{minipage}[t]{0.32\textwidth}
        \centering
        \includegraphics[width=\linewidth]{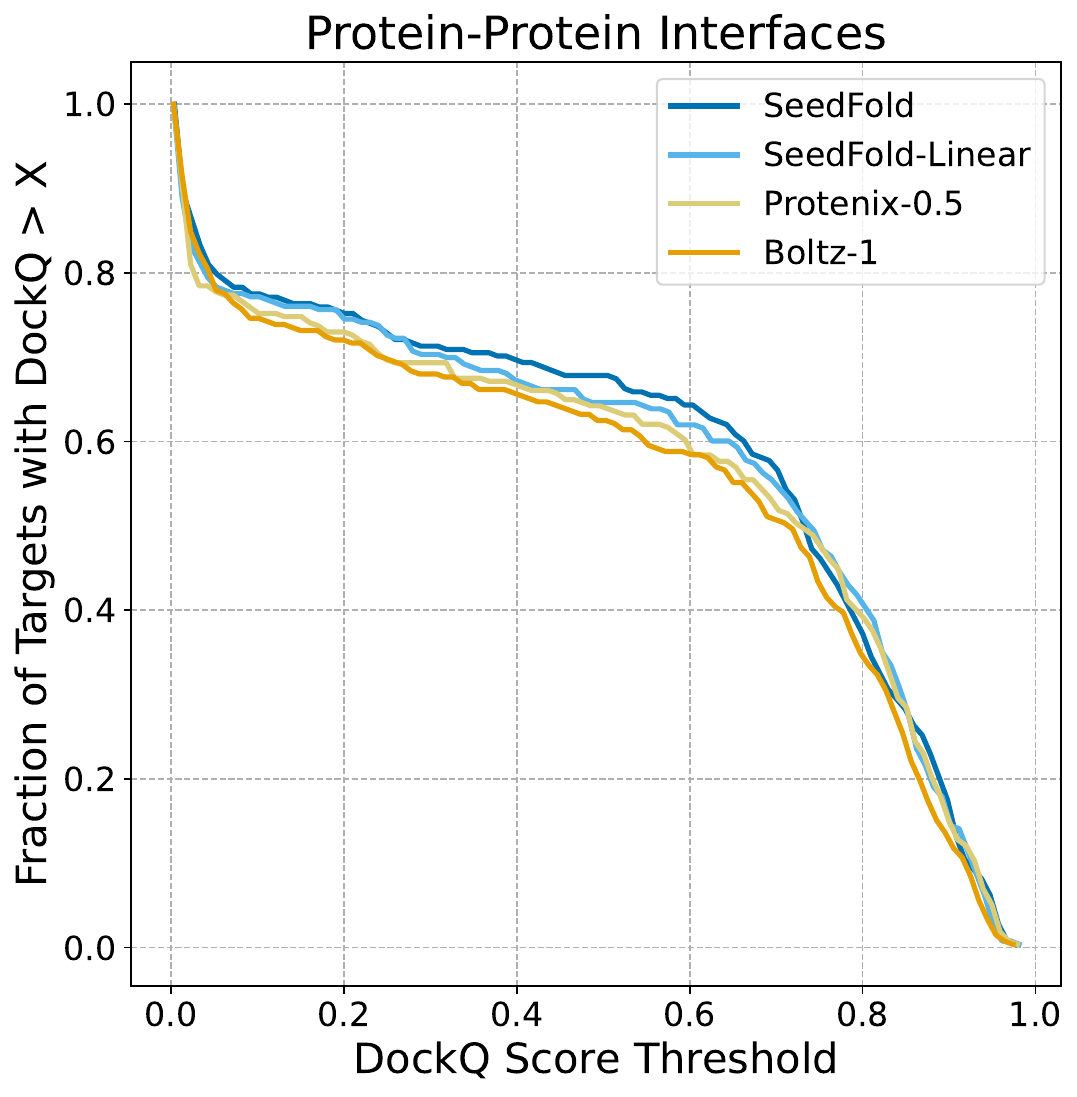}
    \end{minipage}
    \hfill
    \begin{minipage}[t]{0.32\textwidth}
        \centering
        \includegraphics[width=\linewidth]{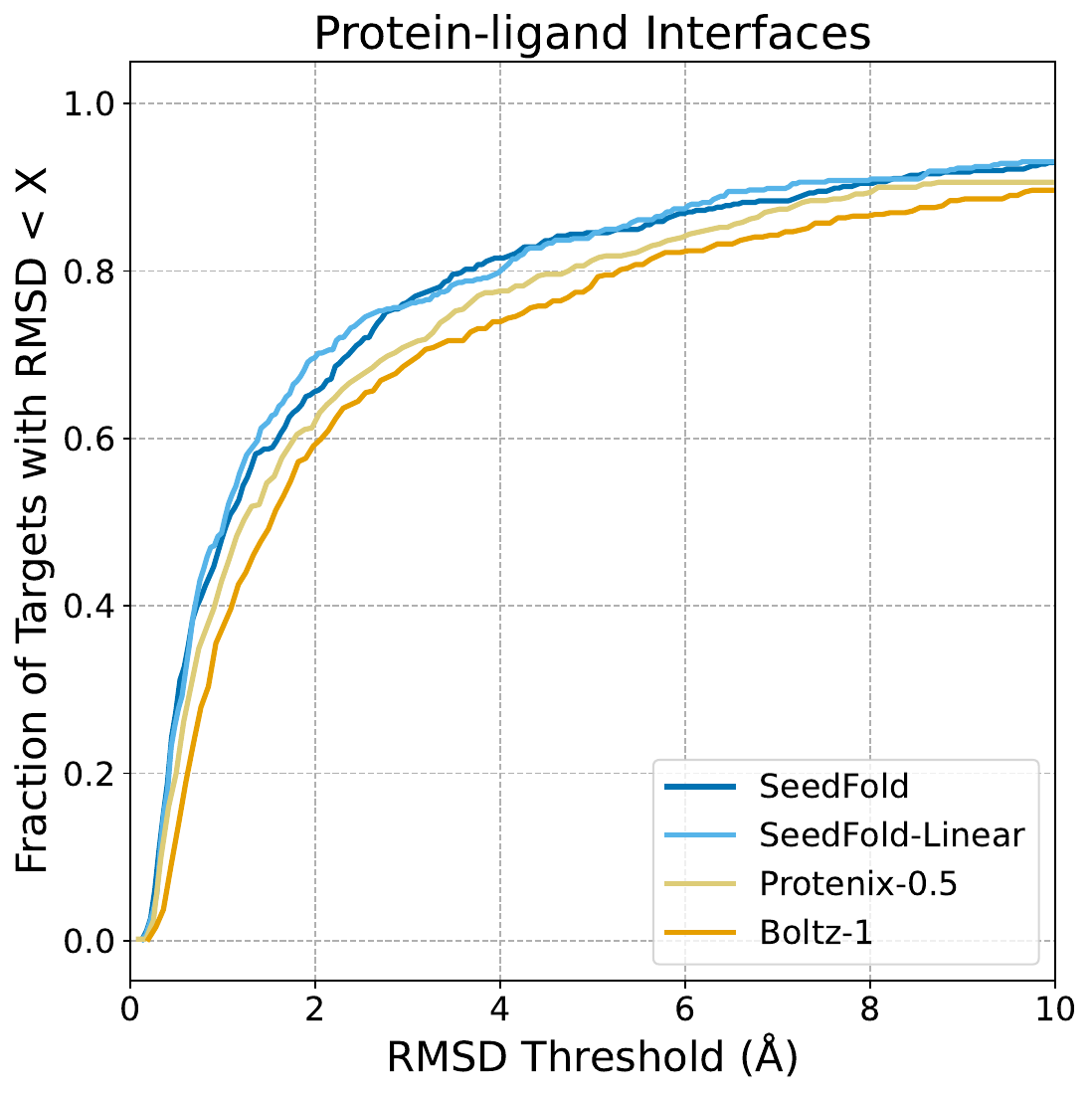}
    \end{minipage}
    \caption{Cumulative distribution of interface prediction success rates across different modalities. We compare \texttt{SeedFold} and \texttt{SeedFold-Linear} against Boltz-1~\cite{wohlwend2025boltz} and Protenix-0.5~\cite{bytedance2025protenix}. Note that AlphaFold 3's detailed distribution metrics are not available due to license restrictions; their performance could be found in FoldBench~\cite{xu2025benchmarking}.}
    \label{fig:interface_distribution}
\end{figure}


\begin{figure}[t]
    \centering
    \includegraphics[width=1.0\linewidth]{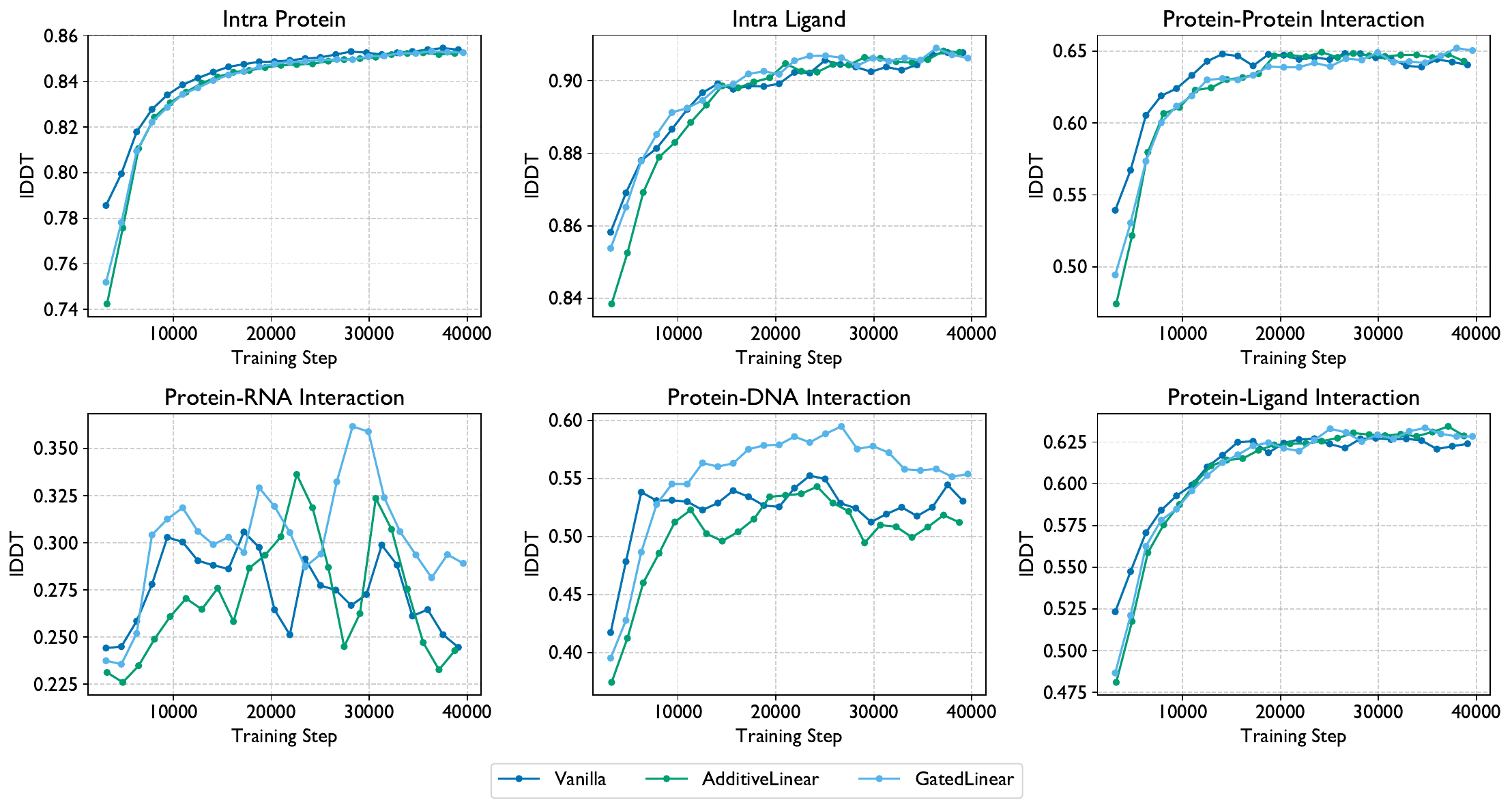}
    \caption{The validation scores across various metrics for different types of triangular attention are presented as follows. Linear triangular attention achieves on-par results on most metrics; moreover, in RNA/DNA-related tasks, \texttt{GatedLinearTriAtt} outperforms \texttt{AdditiveLinearTriAtt} and the standard attention mechanism.}
    \label{fig:triatt_ablation}
\end{figure}

\paragraph{Overall Peformance} As detailed in Table~\ref{tab:main_results}, our models demonstrate state-of-the-art or highly competitive performance across the majority of FoldBench tasks. The vanilla \texttt{SeedFold} model establishes new benchmarks in protein monomer prediction (lDDT $0.8889$), antibody-antigen interfaces ($53.21\%$ DockQ), and protein-RNA interfaces ($65.31\%$ DockQ), notably outperforming AlphaFold 3. Concurrently, the \texttt{SeedFold-Linear} variant shows its strength by leading in protein-ligand prediction ($66.48\%$ success rate) and protein-protein interfaces ($74.14\%$ DockQ). In the remaining tasks, including protein-DNA and nucleic acid monomer prediction, both models deliver strong results that are competitive with AlphaFold 3 and consistently surpass other open-source methods. Collectively, these results validate that our width-scaling strategy effectively enhances model capacity and that the linear triangular attention offers a computationally efficient yet powerful alternative to standard attention mechanisms.

\paragraph{Interface Prediction}
To further dissect the performance on key interface prediction tasks, we present the cumulative distribution of success rates in Figure~\ref{fig:interface_distribution}. These plots illustrate the fraction of targets achieving a quality score above a given threshold, offering a more granular view than summary metrics alone. For antibody-antigen interfaces, \texttt{SeedFold} demonstrates a commanding lead over all other models across the entire DockQ score range, indicating its superior accuracy and reliability for this challenging task. In protein-protein interface prediction, while all models perform competitively, both \texttt{SeedFold} and \texttt{SeedFold-Linear} maintain a consistent advantage over Protenix-0.5 and Boltz-1. For protein-ligand interfaces, our models again show a clear superiority, with \texttt{SeedFold-Linear} achieving the best performance. These detailed distributions underscore the significant improvements, providing more robust and accurate predictions across critical interface modeling domains.






\subsection{Ablation Studies}

We conduct ablation studies to understand the contribution of each component.


\paragraph{Attention Mechanism}
To validate our design choices of different attention mechanisms, we conducted a comprehensive ablation study comparing three attention variants: (i) the standard softmax attention (\texttt{VanillaTriAtt}), (ii) a linear version with an additive feature map (\texttt{AdditiveLinearTriAtt}), and (iii) our proposed gated version (\texttt{GatedLinearTriAtt}). As shown in Figure~\ref{fig:triatt_ablation}, the validation curves reveal that both linear attention variants achieve performance on par with the vanilla attention across most prediction tasks, including intra-protein and protein-protein interactions. Notably, for tasks involving nucleic acids (protein-RNA and protein-DNA), the \texttt{GatedLinearTriAtt} variant demonstrates a clear and consistent advantage over both the additive version and the standard softmax attention, indicating its enhanced capability in handling diverse molecular types.

\begin{figure}[h]
    \centering
    \begin{minipage}[t]{0.32\textwidth}
        \centering
        \includegraphics[width=\linewidth]{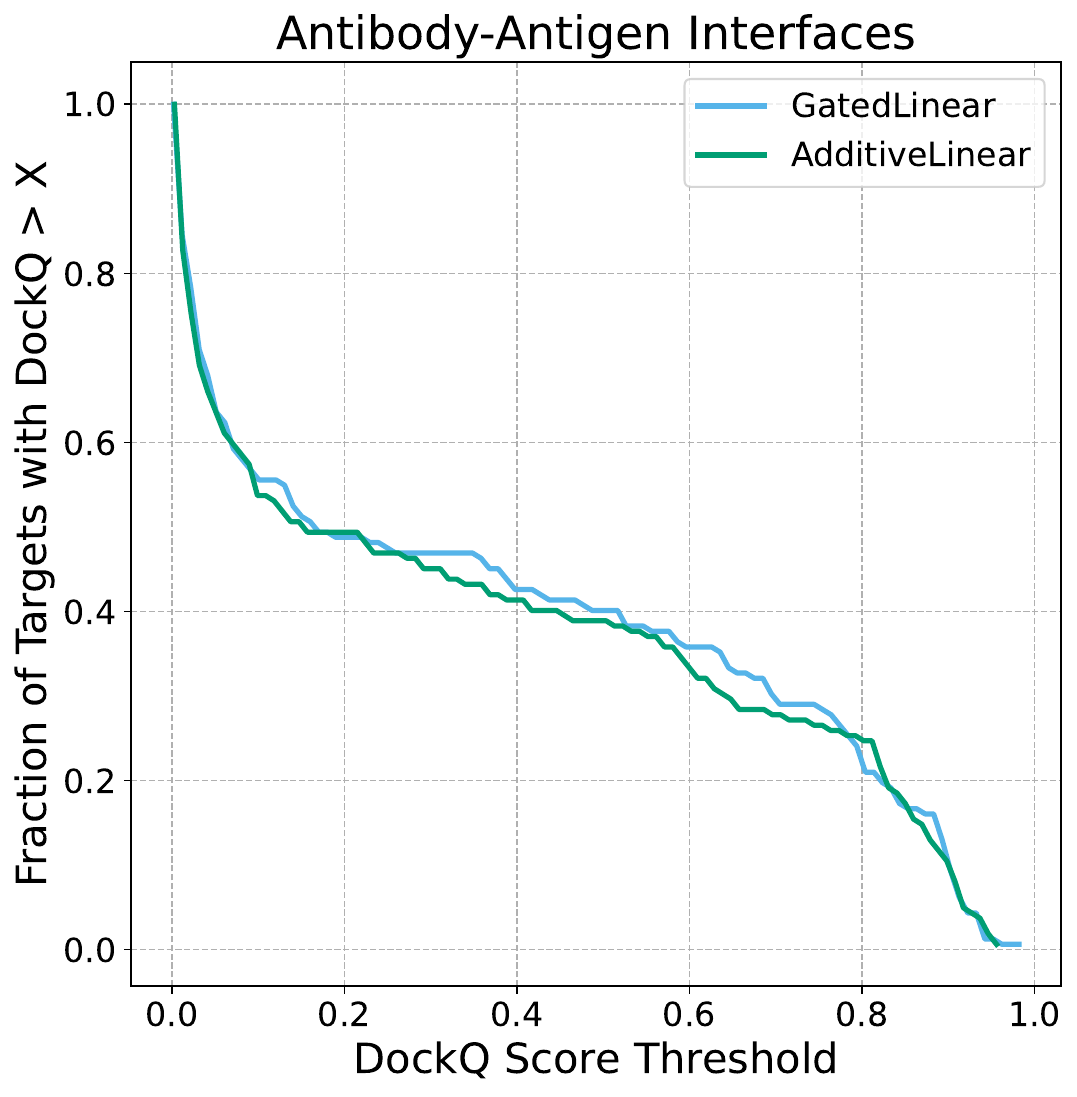}
    \end{minipage}
    \hfill
    \begin{minipage}[t]{0.32\textwidth}
        \centering
        \includegraphics[width=\linewidth]{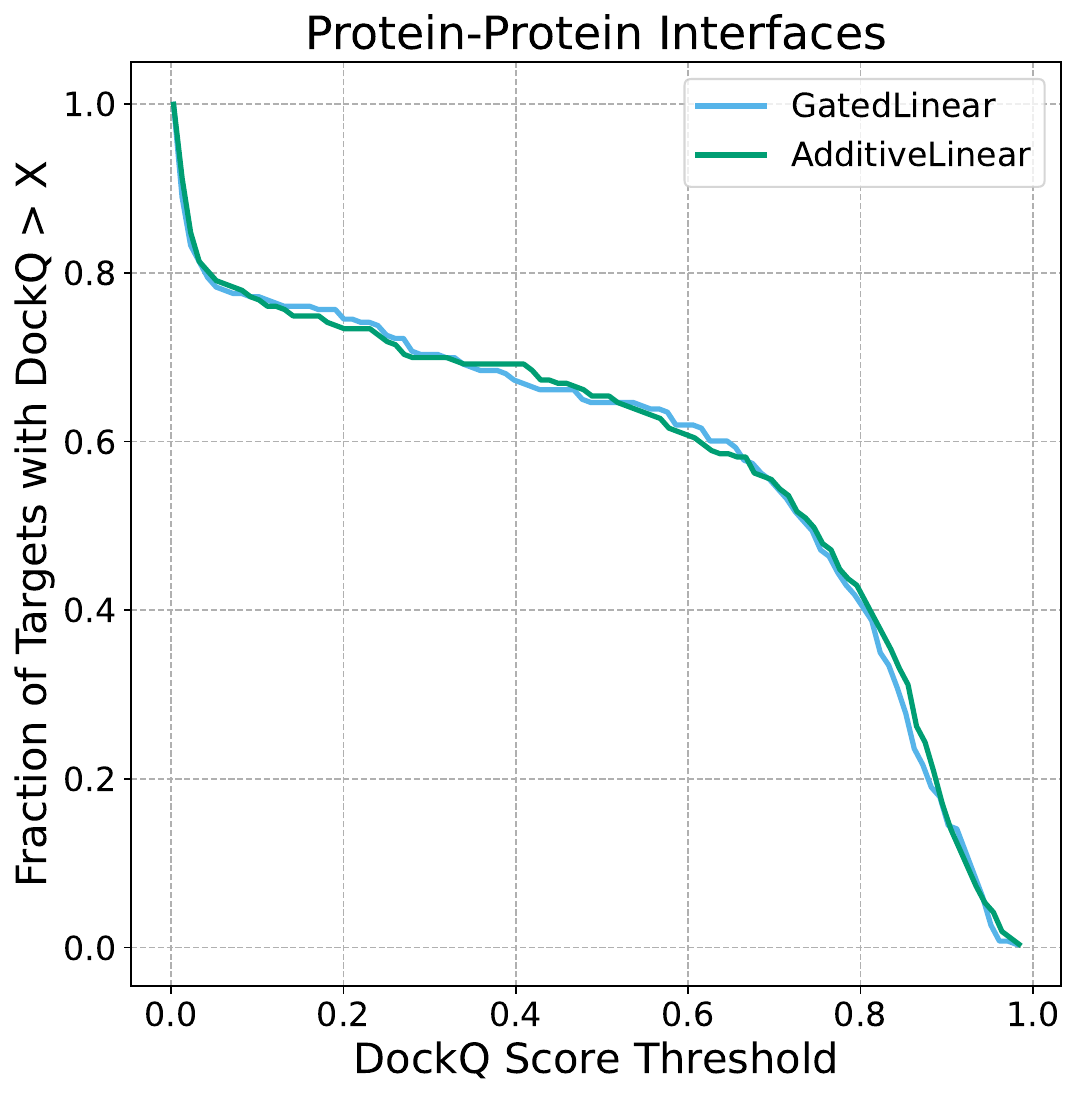}
    \end{minipage}
    \hfill
    \begin{minipage}[t]{0.32\textwidth}
        \centering
        \includegraphics[width=\linewidth]{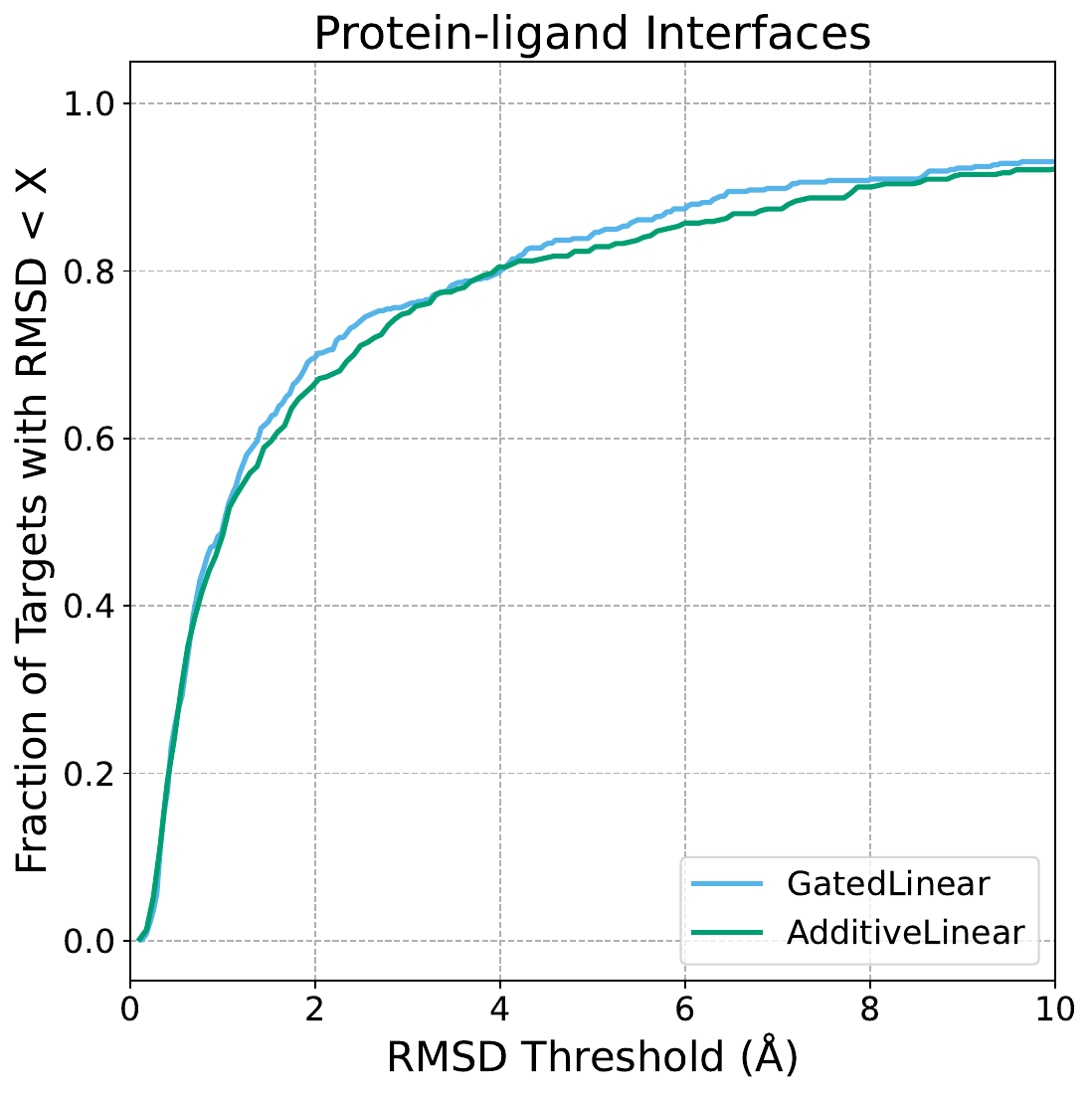}
    \end{minipage}
    \caption{Cumulative distribution of interface prediction success rates across different modalities for linear attention based models. We find that \texttt{GatedLinearTriAtt} slightly outperforms the \texttt{AdditiveLinearTriAtt} module on antibody-antigen and protein-ligand interaction. Therefore, we finally chose \texttt{GatedLinearTriAtt} as the default attention configuration.}
    \label{fig:lin_interface_distribution}
\end{figure}

To further refine our choice, we performed a detailed head-to-head comparison between \texttt{GatedLinearTriAtt} and \texttt{AdditiveLinearTriAtt} on key interface prediction tasks (Figure~\ref{fig:lin_interface_distribution}). The cumulative distribution plots show that \texttt{GatedLinearTriAtt} holds a slight but consistent edge in antibody-antigen and protein-ligand interface predictions. Given its superior performance on nucleic acid and antibody tasks without compromising efficacy elsewhere, we selected \texttt{GatedLinearTriAtt} as the default configuration for our \texttt{SeedFold-Linear} model, confirming it as a powerful and efficient alternative to standard attention.

\paragraph{Monomer Distillation}
Monomer distillation data is a key component of our training set, designed to ground the model's understanding of fundamental protein geometry. To verify its importance throughout the entire training process, we conducted an ablation study. Starting from a checkpoint at $47,612$ steps (trained on the full dataset), we continued training under two conditions: one with the complete dataset, and another where the monomer distillation data was removed. Figure~\ref{fig:distill} (in Appendix~\ref{appendix:data}) shows that removing the distillation data leads to a significant and immediate degradation in intra-protein structure prediction accuracy. While the impact on direct interface metrics appears less pronounced in the short term, maintaining the model's fundamental ability to accurately fold single protein chains is critical for its overall predictive power. This finding underscores the importance of including monomer distillation data throughout the entire training process to prevent knowledge decay and ensure the model's robustness.

\section{Conclusion and Future Work}

In this paper, we present a folding model, \texttt{SeedFold}, which scales the data size and model size with a scalable attention module. \texttt{SeedFold} achieves state-of-the-art performance on FoldBench. Despite the scaling strategies proposed in this study, we leave two additional directions for future work:



\paragraph{Mixture of Experts} There are two primary motivations for adopting mixture-of-experts (MoE) techniques in folding models. Firstly, learning MoEs is significantly more computationally efficient, particularly for architectures that scale cubically with token length. Secondly, a challenge in AF3-like folding models is the conflict in gradient updates across multiple tasks; specifically, the objectives of learning nucleic acids, monomers, ligands, and their respective interactions can lead the networks to update in divergent directions.

\paragraph{Post-training Scaling} Diffusion-based folding models tend to exhibit hallucination. We hypothesize that supervised learning on the training set can only provide limited signals for predicting biomolecular structures. Post-training scaling techniques, including reinforcement learning from ``X'' feedback (RLxF) and test-time compute (TTC), may help align the distribution of folding models to be more desirable~\cite{lai2025survey}.

\newpage

\section*{Contributions}

\textbf{Project Lead}

Yi Zhou$^{1*}$, Chan Lu$^{1*}$

$^{*}$Equal contribution. Listing order is random.

\textbf{Contributors}

Yiming Ma$^{1,2,\dagger}$, Wei Qu$^{1}$, Fei Ye$^{1}$, Kexin Zhang$^{1,3,\dagger}$, Lan Wang$^{1}$, Minrui Gui$^{1,4,\dagger}$

$\dagger$ Work is done during their internship at Bytedance Seed.

\textbf{Overall Technical Lead}

Quanquan Gu$^1$

\textbf{Affiliation}

$^1$ByteDance Seed

$^2$Peking University

$^3$ShanghaiTech University

$^4$University of California, Los Angeles

\section*{Acknowledgments}

We thank Liang Hong, Zaixiang Zheng, Xinyou Wang, Jiasheng Ye, Jing Yuan, Yilai Li, Zhenghua Wang, Yuning Shen, Cheng-Yen Hsieh, Huizhuo Yuan, as well as other colleagues at ByteDance for their discussions and support.

\clearpage

\bibliographystyle{unsrt}
\bibliography{main}

\clearpage

\beginappendix

\section{Triton Kernel}
\label{appendix:triton}
\begin{figure}[h]
    \centering
    \includegraphics[width=1.0\linewidth]{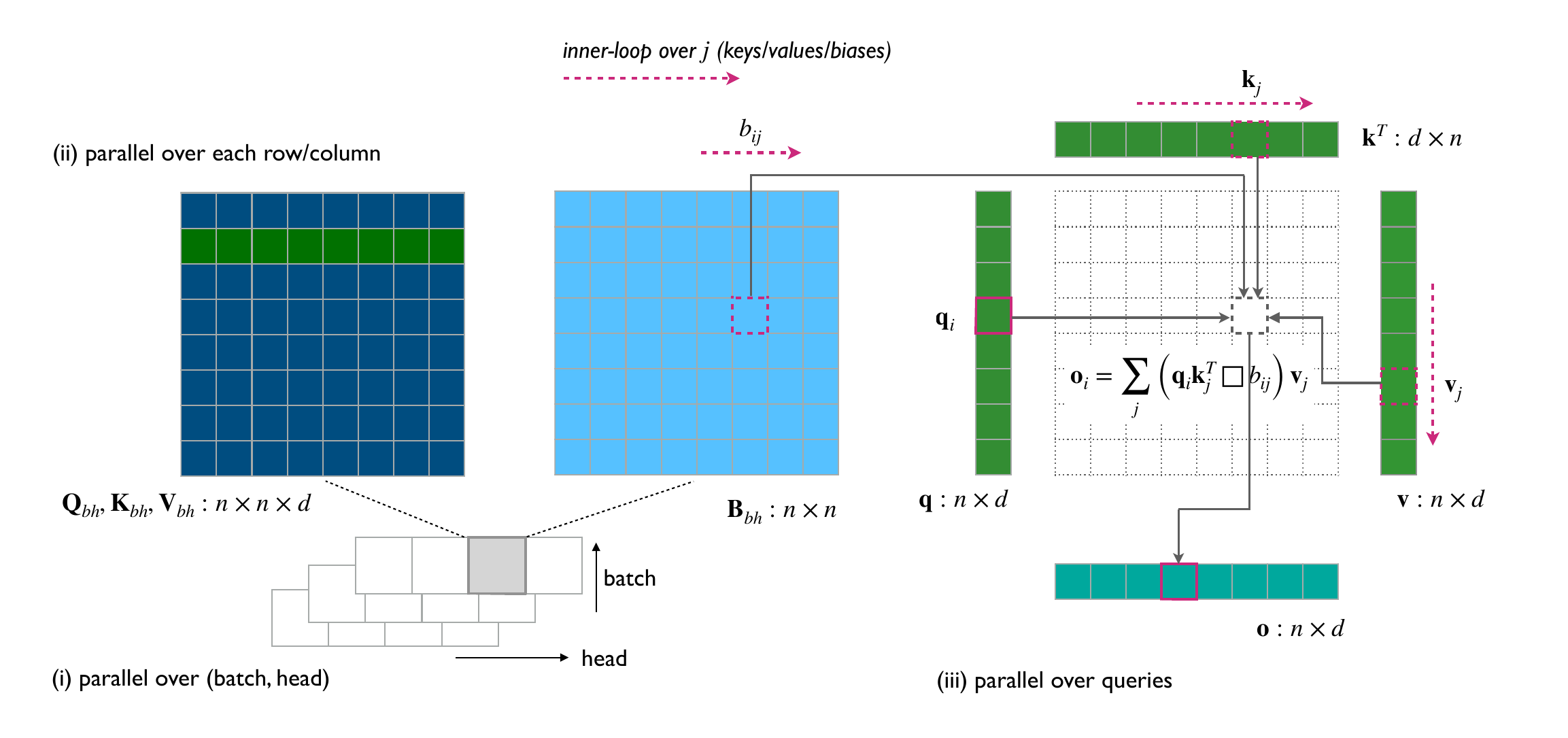}
    \caption{The Triton implementation of \texttt{LinearTriangularAttention}.}
    \label{fig:triton}
\end{figure}

Given $\mathbf{Q},\mathbf{K},\mathbf{V} \in \mathbb{R}^{b\times h\times n \times n \times d}$ and $\mathbf{B} \in \mathbb{R}^{b\times h\times n \times n}$, where $b$, $h$, and $n$ denote the batch size, number of attention heads, and token length, we implemented a Triton~\cite{tillet2019triton} kernel for the $\mathtt{GatedLinearTriangularAttention}$ module:
\begin{align}
\mathbf{O} &= \left(\left(\mathbf{Q}\mathbf{K}^{T}\right)\odot\mathbf{B}\right)\mathbf{V}
\end{align}
Its gradient can be computed by:
\begin{align}
\mathrm{d}\mathbf{V}&=\left(\left(\mathbf{Q}\mathbf{K}^{T}\right)\odot\mathbf{B}\right)^T\cdot\mathrm{d}\mathbf{O} \\
\mathrm{d}\mathbf{Q}&=\left(\left(\mathrm{d}\mathbf{O}\cdot\mathbf{V}^T\right)\odot\mathbf{B}\right)\cdot\mathbf{K} \\
\mathrm{d}\mathbf{K}&=\left(\left(\mathrm{d}\mathbf{O}\cdot\mathbf{V}^T\right)\odot\mathbf{B}\right)^T\cdot\mathbf{Q} \\
\mathrm{d}\mathbf{B}&=\left(\mathrm{d}\mathbf{O}\cdot\mathbf{V}^T\right)\odot\left(\mathbf{Q}\mathbf{K}^T\right)
\end{align}
Notably, the same kernel can be used for computing $\mathbf{O}$, $\mathrm{d}\mathbf{Q}$, $\mathrm{d}\mathbf{K}$, and $\mathrm{d}\mathbf{V}$, whereas $\mathrm{d}\mathbf{B}$ can be implemented through two blocked matrix multiplications. Figure~\ref{fig:triton} illustrates the implementation of this kernel. We treat the $b\times h$ dimensions as a single batch dimension and parallelize over the first dimension of size $n$. Within each process, we iterate over the second dimension of size $n$ to aggregate the result, thereby avoiding the materialization of the full $n\times n$ matrix.

\section{Datasets}
\label{appendix:data}

\begin{figure}[h]
    \centering
    \includegraphics[width=0.4\linewidth]{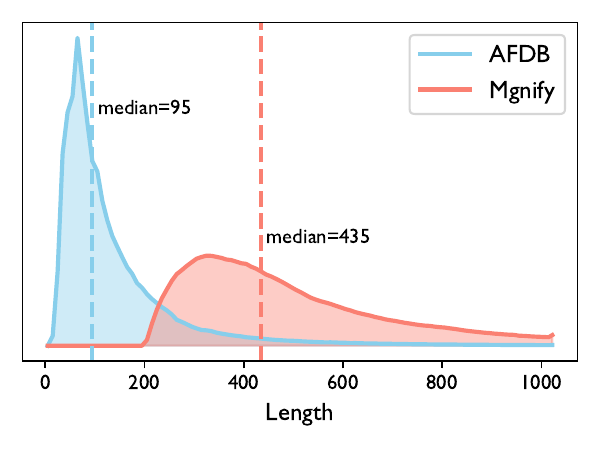}
    \caption{The probability density distribution of sequence lengths for different datasets. The AFDB dataset is dominated by short sequences, while MGnify contains more long sequences.}
    \label{fig:data_length_dist}
\end{figure}

We plot the probability density distribution of the protein lengths in the AFDB dataset and the MGnify dataset in Figure~\ref{fig:data_length_dist}. The length distribution of AFDB is skewed toward short proteins, with a median length of $95$, whereas the length distribution of MGnify is more uniform. 

\begin{figure}[h]
    \centering
    \includegraphics[width=1.0\linewidth]{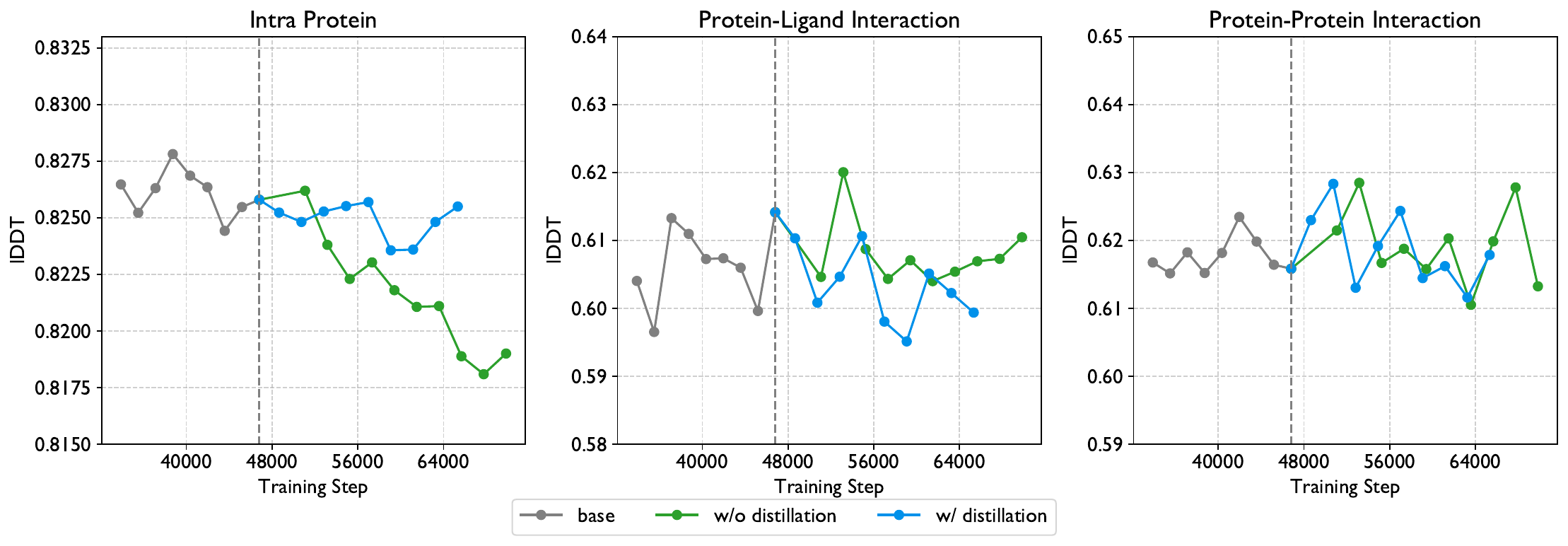}
    \caption{The effect of removing distillation data at certain stages. The grey curve represents the base model from which the continual training stage initiates. The green and blue curves denote the model trained with and without distillation data, respectively. Removal of distillation data leads to a significant degradation in intra-protein lDDT.}
    \label{fig:distill}
\end{figure}

\section{Training Stability during Scaling}
\label{appendix:scaling_stability}

Scaling model capacity introduced several training stability issues. This section documents the challenges and our solutions. As the model size increases, we observed gradient explosion issues, particularly in the early stages of training. To address this, we applied gradient clipping and adjusted the learning rate schedule.


\begin{itemize}
    \item \textbf{Learning Rate Adjustment}: When scaling from Medium ($256$-width) to Large ($512$-width), we found it necessary to reduce the learning rate and adjust optimization hyperparameters to stabilize training.
    \item \textbf{Warm-up Strategy}: We employed an extended warm-up period from $1000$ to $3000$ for larger models to ensure stable convergence in the initial training phase.
\end{itemize}

\begin{figure}[!htbp]
    \centering
    \includegraphics[width=0.8\linewidth]{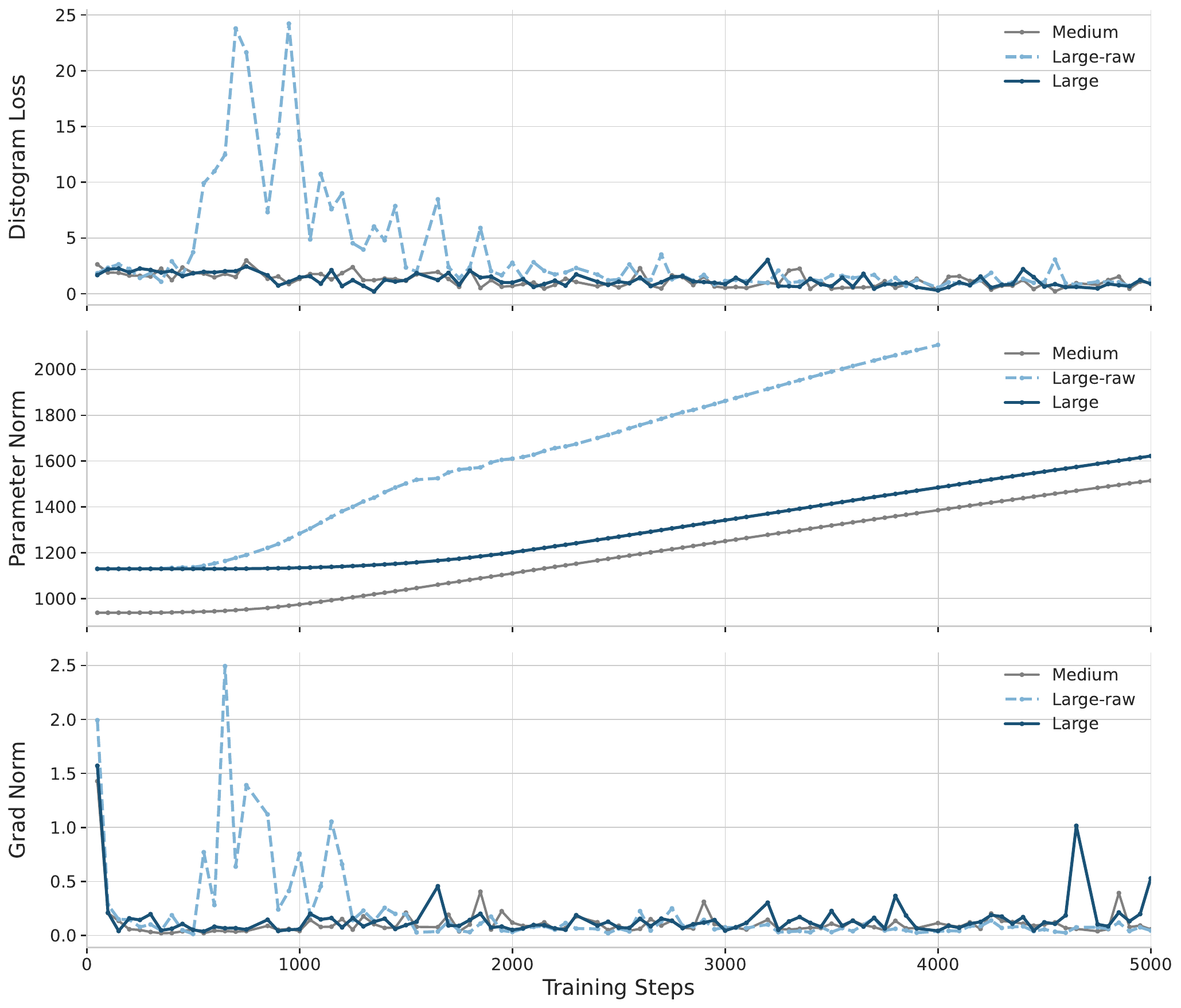}
    \caption{Training dynamics during the early $5000$ steps. The \texttt{Large-Raw} model shows unstable training with loss spikes, while the \texttt{Large} model with optimized hyperparameters exhibits stable convergence.}
    \label{fig:scaling_stability_training}
\end{figure}

\begin{figure}[!htbp]
    \centering
    \includegraphics[width=0.8\linewidth]{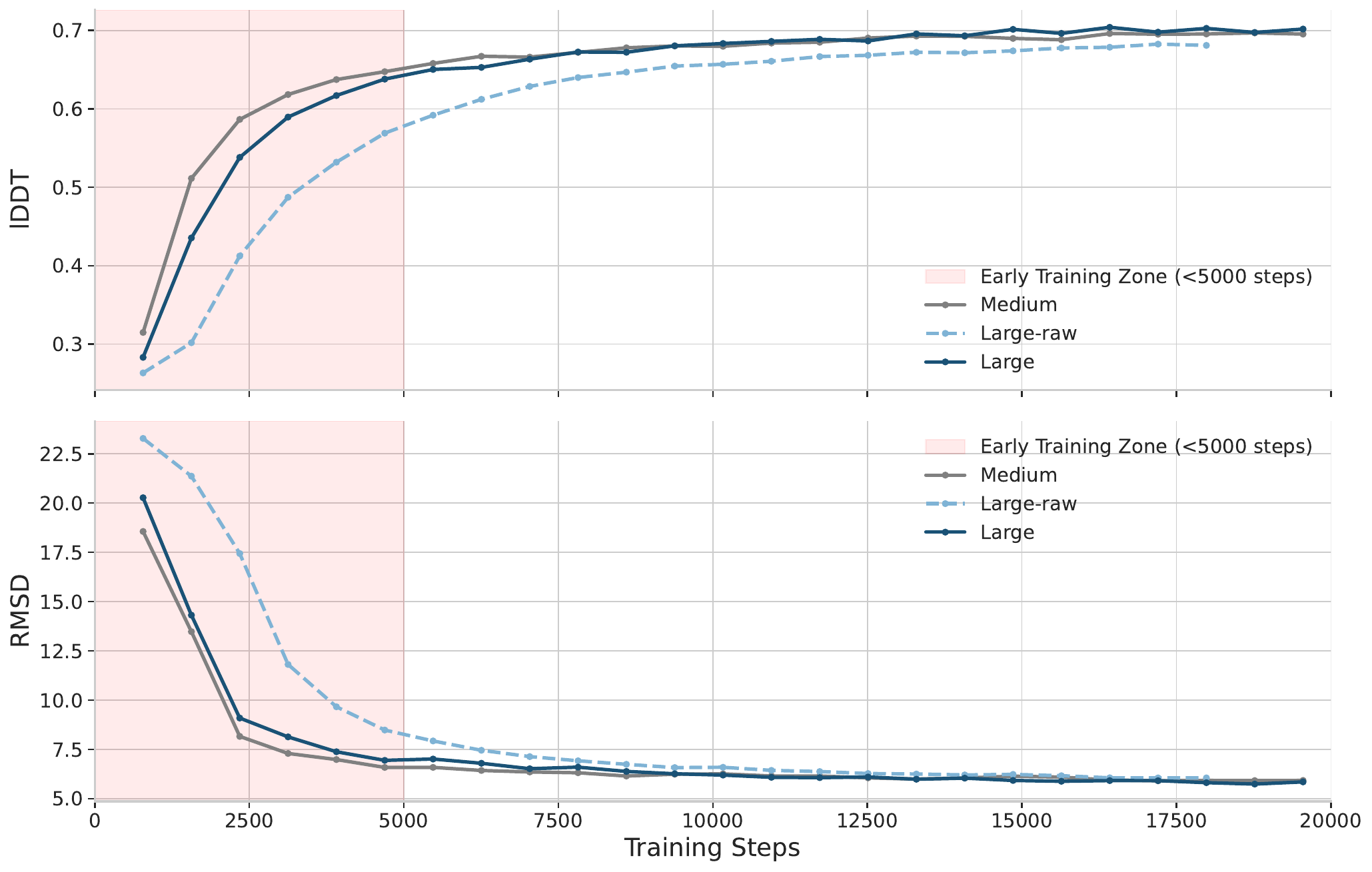}
    \caption{Validation metrics comparison between \texttt{Large-Raw} and \texttt{Large}. The stability optimization leads to improved performance across validation metrics.}
    \label{fig:scaling_stability_validation}
\end{figure}

We denote the original Large model without stability optimization (smaller learning rate with longer warmup steps) as \texttt{Large-Raw}, and the optimized version as \texttt{Large}. Figure~\ref{fig:scaling_stability_training} illustrates the training dynamics during the early $5000$ steps. The \texttt{Large-Raw} model exhibits unstable training behavior with significant loss fluctuations, while the \texttt{Large} model with optimized hyperparameters demonstrates stable convergence. Figure~\ref{fig:scaling_stability_validation} presents the validation metrics comparison, showing that the stability optimization not only improves training dynamics but also leads to better final performance.

\section{Design Space of Linear Triangular Attention}

Although modern linear attention has been thoroughly studied in autoregressive language models~\cite{kimiteam2025kimilinearexpressiveefficient,minimax2025minimax01scalingfoundationmodels}, it remains an open question as to how one should incorporate the bias term into non-autoregressive linear attention, i.e., linear triangular attention. This section briefly discusses the design space of linear triangular attention. 

\begin{figure}[h]
    \centering
    \includegraphics[width=1.0\linewidth]{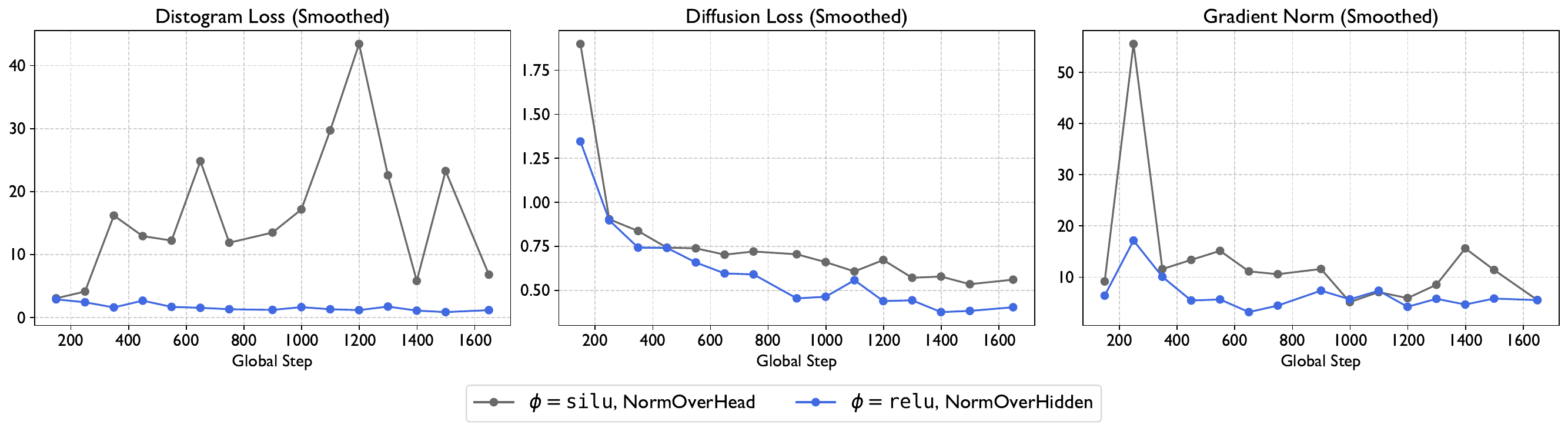}
    \caption{Our choice of feature function and the normalization strategy stablizes the training process, while the original version fails to converge. The distogram loss serves as a direct signal to indicate the convergence.}
    \label{fig:linatt_choices}
\end{figure}

Despite the simplicity of the mathematical formulation of linear attention, several design choices remain to be determined in practical implementation. We start with the basic form, $\phi(\mathbf{Q}_i)\phi(\mathbf{K}_i^T)\mathbf{V}_i$, where $\mathbf{Q}_i, \mathbf{K}_i, \mathbf{V}_i\in\mathbf{R}^{n\times d}$ are the $i$-th row of the queries, keys, and values. The first question is how to bound the values of output feature vectors to prevent the explosion or vanishing of hidden states. A simple yet effective choice is to normalize the attention score with their summation, similar to the \texttt{softmax} attention:
\begin{align}
\frac{\left[\phi(\mathbf{Q}_i)\phi(\mathbf{K}_i^T)\Box\psi(\mathbf{B})\right]\mathbf{V}_i}{\left[\phi(\mathbf{Q}_i)\phi(\mathbf{K}_i^T)\Box\psi(\mathbf{B})\right]\mathbf{1}^{n\times 1}},
\end{align}
where $\mathbf{B}\in\mathbb{R}^{n\times n}$, and $\Box$ can be an arbitrary operation. Such a formulation has proven its effectiveness in modern generative tasks, such as ViT-based high-resolution image synthesis~\cite{xie2023efficient,xie2024sana}. However, there is a potential risk for applying this architecture: if we want to introduce a bias term into the formulation, we still have to materialize a $n\times n$ matrix in the denominator, which is memory-intensive. Besides, a challenge of training linear attention in this form is that the unbounded gradients may lead to unstable convergence~\cite{qin2022devil}. We adopt an alternative approach proposed in Lightning Attention~\cite{qin2024lightning}: dropping the normalization term in the denominator while applying a \texttt{LayerNorm} layer to the output:
\begin{align}
\mathtt{LayerNorm}_{\texttt{over-head/hidden}}\left(\left[\phi(\mathbf{Q}_i)\phi(\mathbf{K}_i^T)\Box\psi(\mathbf{B})\right]\mathbf{V}_i\right).
\end{align}
In the above formulation, there are two final options to be determined: (i) how to select the feature function $\phi$, and (ii) whether to apply the normalization layer to the hidden dimension or the head dimension. Through comprehensive ablation studies, we found that the original configuration ($\phi=\mathtt{silu}$, $\texttt{LayerNormOverHead}$) in Lightning Attention~\cite{qin2024lightning} failed to converge in folding tasks, so we switched to ($\phi=\mathtt{relu}$, $\texttt{LayerNormOverHidden}$), which enabled us to achieve substantially better performance. In Figure~\ref{fig:linatt_choices}, we employ the distogram loss as an indicator to evaluate how well the module fits the data. 

\section{Comparison of Two Linear Attention}

\end{document}